\documentclass[sigconf,authorversion,nonacm]{acmart}

\usepackage[T1]{fontenc}
\usepackage{xspace}
\usepackage{graphicx}
\usepackage{xcolor}
\usepackage{subcaption}
\usepackage{balance}
\usepackage{booktabs}
\usepackage{tabularx}
\usepackage{diagbox}
\usepackage[utf8]{inputenc}
\usepackage{makecell}
\usepackage{enumitem}

\usepackage{cleveref}

\newcommand{\sys}{\mbox{\textsc{LST-Bench}}\xspace}
\newcommand{\company}{Microsoft\xspace}

\newcommand{\lst}{\textsc{LST}\xspace}
\newcommand{\lsts}{\textsc{LSTs}\xspace}

\newcommand{\cut}[1]{}

\newcommand{\smallsection}[1]{\vspace{1mm}\noindent\textbf{#1.}}	%

\newcommand{\rulesep}{\unskip\ \vrule\ }

\newcommand{\myparagraph}[1]{\vspace{1mm}\noindent\textbf{#1}}

\newcommand{\workload}[1]{$\textrm{WP}$#1\xspace}
\newcommand{\baseworkload}{$\textrm{W}$0\xspace}
\newcommand{\phase}[1]{{\sc #1}\xspace}
\newcommand{\load}{{\sc Load}\xspace}
\newcommand{\singleuser}{{\sc Single User}\xspace}
\newcommand{\throughput}{{\sc Throughput}\xspace}
\newcommand{\datamaintenance}{{\sc Data Maintenance}\xspace}
\newcommand{\optimize}{{\sc Optimize}\xspace}
\newcommand{\timetravel}{{\sc Time Travel}\xspace}
\newcommand{\ld}[1]{{\sc L}\xspace}
\newcommand{\su}[1]{{\sc SU}\xspace}
\newcommand{\dm}[1]{{\sc DM}\xspace}
\newcommand{\opt}[1]{{\sc O}\xspace}
\newcommand{\sutt}[1]{{\sc SU-TT}\xspace}

\usepackage[framemethod=TikZ]{mdframed}
\newcounter{insight}[section]
\renewcommand{\theinsight}{\arabic{insight}}

\usepackage{threeparttable}

\usepackage[framemethod=TikZ]{mdframed}
\newcounter{observation}[section]
\renewcommand{\theobservation}{\arabic{observation}}

\AtBeginDocument{%
  \providecommand\BibTeX{{%
    \normalfont B\kern-0.5em{\scshape i\kern-0.25em b}\kern-0.8em\TeX}}}

\usepackage[a-2b]{pdfx}

\begin{document}
\title{\sys: Benchmarking Log-Structured Tables in the Cloud}

\author{Jes\'us Camacho-Rodr\'iguez}
\email{jesusca@microsoft.com}
\orcid{0009-0008-9151-6024}
\affiliation{%
  \institution{Microsoft}
  \country{USA}
}
\author{Ashvin Agrawal}
\email{ashvin.agrawal@microsoft.com}
\orcid{0009-0004-7862-0995}
\affiliation{%
  \institution{Microsoft}
  \country{USA}
}
\author{Anja Gruenheid}
\email{anja.gruenheid@microsoft.com}
\orcid{0009-0009-2547-8610}
\affiliation{%
  \institution{Microsoft}
  \country{Switzerland}
}
\author{Ashit Gosalia}
\email{ashit.gosalia@microsoft.com}
\orcid{0009-0003-7939-6692}
\affiliation{%
  \institution{Microsoft}
  \country{USA}
}
\author{Cristian Petculescu}
\email{cristp@microsoft.com}
\orcid{0009-0006-9007-4733}
\affiliation{%
  \institution{Microsoft}
  \country{USA}
}
\author{Josep Aguilar-Saborit}
\email{jaguilar@microsoft.com}
\orcid{0009-0009-4641-0236}
\affiliation{%
  \institution{Microsoft}
  \country{USA}
}
\author{Avrilia Floratou}
\email{avrilia.floratou@microsoft.com}
\orcid{0009-0007-5760-8657}
\affiliation{%
  \institution{Microsoft}
  \country{USA}
}
\author{Carlo Curino}
\email{carlo.curino@microsoft.com}
\orcid{0000-0003-3712-7358}
\affiliation{%
  \institution{Microsoft}
  \country{USA}
}
\author{Raghu Ramakrishnan}
\email{raghu@microsoft.com}
\orcid{0009-0007-5086-7664}
\affiliation{%
  \institution{Microsoft}
  \country{USA}
}

\renewcommand{\shortauthors}{Jes\'us Camacho-Rodr\'iguez et al.}

\begin{abstract}
Data processing engines increasingly leverage distributed file systems for scalable, cost-effective storage. 
While the Apache Parquet columnar format has become a popular choice for data storage and retrieval, the immutability of Parquet files renders it impractical to meet the demands of frequent updates in contemporary analytical workloads. 
Log-Structured Tables (\lsts), such as Delta Lake, Apache Iceberg, and Apache Hudi, offer an alternative for scenarios requiring data mutability, providing a balance between efficient updates and the benefits of columnar storage. 
They provide features like transactions, time-travel, and schema evolution, enhancing usability and enabling access from multiple engines. 
Moreover, engines like Apache Spark and Trino can be configured to leverage the optimizations and controls offered by \lsts to meet specific business needs. 
Conventional benchmarks and tools are inadequate for evaluating the transformative changes in the storage layer resulting from these advancements, as they do not allow us to measure the impact of design and optimization choices in this new setting. 

In this paper, we propose a novel benchmarking approach and metrics that build upon existing benchmarks, aiming to systematically assess \lsts. 
We develop a framework, \sys, which facilitates effective exploration and evaluation of the collaborative functioning of \lsts and data processing engines through tailored {\em benchmark packages}. 
A package is a mix of use patterns reflecting a target workload; \sys makes it easy to define a wide range of use patterns and combine them into a package, and we include a baseline package for completeness. 
Our assessment demonstrates the effectiveness of our framework and benchmark packages in extracting valuable insights across diverse environments. 
The code for \sys is open source and is available at \url{https://github.com/microsoft/lst-bench/}.
\end{abstract}

\maketitle

\section{Introduction}
\label{sec:intro}

Parquet~\cite{parquet} and ORC~\cite{orc} are widely popular columnar file formats designed to optimize data storage and retrieval, with a bias toward read-heavy workloads. 
These files are designed to be immutable: once created, they are read-only and optimized for efficient columnar reads. 
Modern analytical workloads, however, require frequent incremental updates to structured data, i.e., tables, in small batches in order to expedite insights and maximize the business value derived from data. 
Log-Structured Tables (\lsts) effectively cater to this requirement while leveraging the inherent strengths of columnar formats. 
They have become the industry standard and pervasively adopted in the field. 
Several implementations of \lsts have emerged, with Delta Lake~\cite{delta-lake,DBLP:journals/pvldb/ArmbrustDPXZ0YM20}, Apache Iceberg~\cite{apache-iceberg}, and Apache Hudi~\cite{apache-hudi} being the most widely adopted ones.
These \lsts add a metadata layer on top of immutable columnar files to represent versions of tables and specify how data processing engines and applications interact with them\footnote{While these implementations do not yet cover security and access control, we believe that they need to do so; in practice, most systems using them add a security layer.}.

\lsts represent a significant paradigm shift in the storage layer design from traditional warehouse systems. 
Unlike traditional systems that manage their own storage~\cite{exadata,teradata,redshift}, \lsts rely on non-POSIX APIs provided by object stores~\cite{aws-s3,azure-adls,google-cloud-storage,ozone} to enable their features, which can be shared across compute engines. 
\lsts use column-oriented log-structured immutable files instead of the row-oriented in-place updated page files used by traditional OLTP and warehouse database systems~\cite{DBS-024}. 
\lsts are designed specifically for processing large-scale data that receives continuous trickle updates~\cite{iceberg-cdc}, which, while not as frequent as in OLTP-style workloads, differ markedly from the infrequent, large-scale batch updates that traditional warehouses doing in-place updates are designed for~\cite{datawarehousekimball2013,westerman2001data}. 
They provide single-table ACID transactions\footnote{Multi-table transactions are a notable gap in comparison to traditional database systems; we expect this will also be addressed in future.} using \emph{multi-version concurrency control}~\cite{inmemory-mvcc}, creating a new version of tables by `depositing' a new immutable layer of files containing changes made to the dataset. 
This approach also enables features such as \emph{time travel} queries. 
However, one drawback of having multiple layers of version files is the increase in IO operations required to retrieve data for query processing, which can result in slower query execution~\cite{DBLP:conf/sigmod/Camacho-Rodriguez19}. 
To tackle this issue and achieve desired outcomes, \lsts offer various controls, optimizations, and algorithms, including compaction, file sizing, clustering, and caching.

\subsection{Challenges in Benchmarking \lsts}
\label{sec:intro:prob}

The differences between \lsts and traditional warehouse systems introduce unique new challenges for users to learn how to operate and tune \lsts. Benchmarking is the go-to methodology to learn the characteristics of these new systems. 
However, there has been limited research on developing innovative evaluation mechanisms for \lsts aimed at formalizing the complexities of continuously changing performance in long-running deployments or the effect of concurrently running (and often expensive) maintenance operations on files stored on object stores.

As a result, users have been forced to rely solely on previously established benchmarks. 
Current evaluations typically rely on TPC-DS~\cite{DBLP:conf/vldb/OthayothP06}, which has long been the standard OLAP benchmark, and involve running a limited number of queries or using handcrafted queries to test a variety of operations~\cite{brooklyn-data,lhbench}.
We observe that these evaluations suffer from two main limitations that restrict their ability to provide useful insights for \lsts. 
Firstly, they fail to uncover hidden characteristics inherent in \lsts and data processing engines that are crucial in real-world usage scenarios. 
Secondly, they lack specific evaluation metrics that are important for \lsts, such as performance degradation over time. 
Our goal is to propose a framework that complements a base workload such as TPC-DS to address these limitations.

\begin{figure}[t]
    \centering
    \includegraphics[width=\columnwidth]{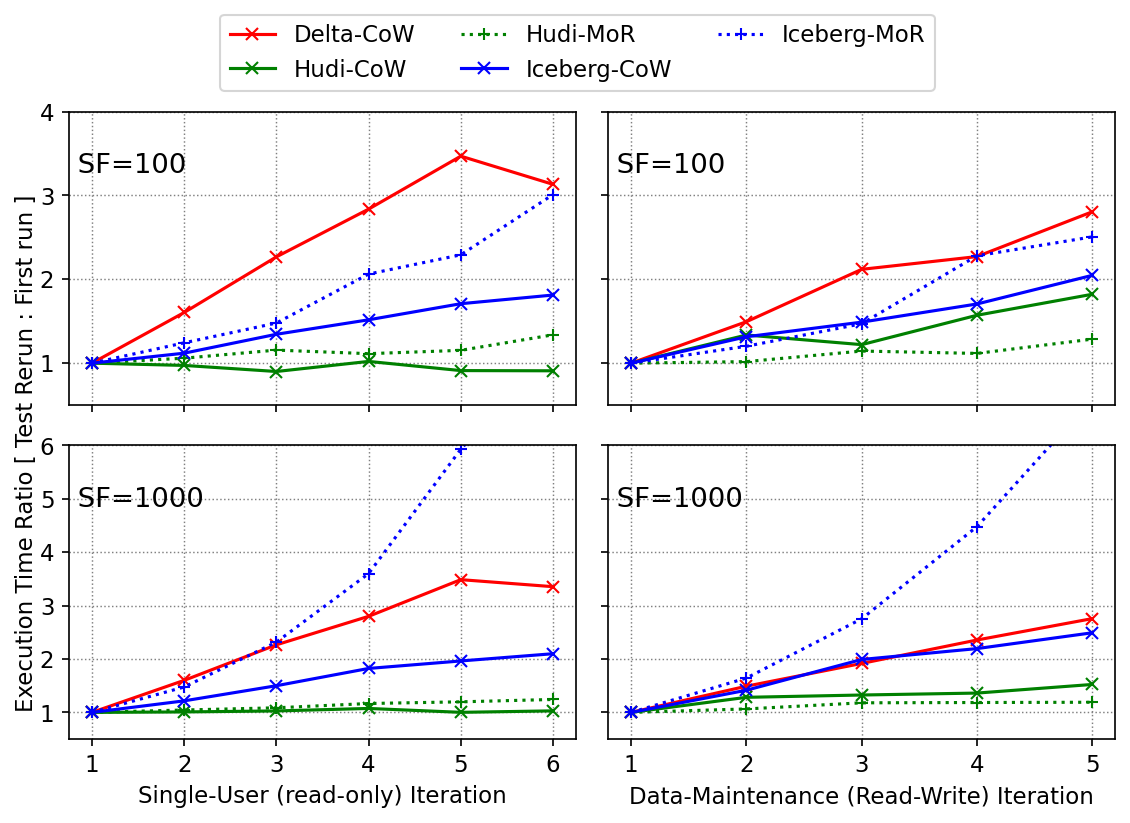}
    \caption{Execution time comparison of TPC-DS \emph{single user} \& \emph{data maintenance} test iterations at different scale factors (SF=100, SF=1000) using various \lsts and strategies such as Copy-on-Write (CoW) and Merge-on-Read (MoR) on Spark.}
    \label{fig:w1_perf_degradation}
\end{figure}

\smallsection{Evaluation Scenarios}
Current benchmarking workloads, such as TPC-DS, reflect a generalized understanding of OLAP tasks and do not consider characteristics such as ($i$)~\emph{longevity}, which involves handling frequent data modifications over a long period of time; ($ii$)~\emph{resilience}, which involves handling multiple data modifications of varying sizes in a regularly optimized table; ($iii$)~\emph{read/write concurrency}, which involves handling multiple sessions reading and writing data simultaneously, potentially using multiple compute clusters; or ($iv$)~\emph{time travel}, which involves querying data at different points in time. 
However, these characteristics are crucial for \lsts, which vary from traditional warehousing systems in building on immutable files and relying heavily on versions and mechanisms for compaction, version control, etc., and also in how they are deployed by customers interested in new features such as time travel.

\begin{example}\label{ex:perf_degradation_w1}
As we mentioned previously, performance degradation due to accumulation of version files over time is an important evaluation factor for \lsts. 
The standard TPC-DS workload involves two rounds of interleaved read and write queries, typically reported together~\cite{DBLP:conf/vldb/OthayothP06}. 
Instead, we conducted experiments with an increased number of TPC-DS \emph{single user} (read-only) query iterations, alongside \emph{data maintenance} (read-write) steps between individual \emph{single user} iterations. 
\Cref{fig:w1_perf_degradation} depicts the results, demonstrating consistent execution time deterioration for both \emph{single user} and \emph{data maintenance} tests (details in \S\ref{sec:6.1_w1}). 
More importantly, by analyzing the results, we can observe interesting trends beyond the second iteration, which would be overlooked if relying solely on the original TPC-DS workload.
\end{example}

\smallsection{Performance Metrics}
Similar to other benchmarks~\cite{tpc-benchmarks}, the TPC-DS specification focuses on a primary metric, namely queries per hour for decision support (QphDS), which provides a single performance measurement that is used for the comparison of systems. 
While this approach simplifies the ranking of multiple systems, it fails to capture essential dimensions that are relevant for evaluating \lsts.
For instance, understanding the performance and efficiency degradation of \lsts over time is crucial to determining how system designers can effectively optimize platforms that rely on \lsts.

\begin{example}\label{ex:qphds_w0}
The results of executing TPC-DS\footnote{\label{qphds}Standard dataset and execution rules were followed, but the audit step was not performed.} (scale factor 1000) with various \lsts and Spark are presented in \Cref{table:qphds_w0}.
However, \emph{QphDS} does not capture the performance degradation of the second run relative to the first (i.e., the increase in latency), depicted in the `Inter-test Degradation' column, which shows that well-performing \lsts can significantly degrade over time.
These numbers indicate that running these \lsts without appropriate mediation will result in low-performance results.
\end{example}

\begin{table}[ht!]
\caption{Latency increase between TPC-DS test iterations.\vspace{-0.5em}}
\resizebox{\columnwidth}{!}{%
\begin{tabular}{ccc}
    \toprule
    \textbf{\lst} & \textbf{Throughput-QphDS} & \textbf{Inter-test Degradation} \\
    \midrule
    Delta & 511K & 2.7 -> 5.2 hrs (\textbf{92\%}) \\
    Hudi-CoW\textsuperscript{\textdagger} & 262K & 6.2 -> 6.5 hrs (\textbf{5\%}) \\
    Hudi-MoR\textsuperscript{\textdaggerdbl} & 112K & 23 -> 24 hrs (\textbf{6\%}) \\
    Iceberg-CoW\textsuperscript{\textdagger} & 549K & 2.7 -> 4 hrs (\textbf{45\%}) \\
    Iceberg-MoR\textsuperscript{\textdaggerdbl} & 493K & 2.9 -> 5 hrs (\textbf{73\%}) \\
    \bottomrule
    \multicolumn{3}{l}{\footnotesize{\textdagger~Copy-on-Write mode \qquad \textdaggerdbl~Merge-on-Read mode}}
\end{tabular}
}
\label{table:qphds_w0}
\end{table}

\smallsection{Framework Flexibility} 
The overall performance and cost of \lsts are significantly influenced by the query engine and algorithmic components that orchestrate optimization tasks such as file clustering, small file compaction, caching, and more. 
For instance, in \S\ref{sec:6.1_w1}, we show that read queries exhibit up to an 85\% improvement in execution time when running on Trino compared to Spark, for both Delta and Iceberg. 
This difference stems mainly from Spark's default distribution mode, which leads to generation of a large number of small data files. 
Consequently, it is important that a benchmarking framework for \lsts enables detection and analysis of such variations. 
In the aforementioned case, we draw our conclusion by correlating the benchmark run with telemetry collected from the cloud storage infrastructure using our framework. 
Moreover, as \lsts are a new proposal with continuously expanding use cases, it is critical that the framework supports extension to new engines,  datasets, and scenarios that go beyond traditional OLAP tasks.

\subsection{Contributions}

This paper presents a benchmarking framework to evaluate open-source \lsts and overcome the limitations of existing benchmarks. 
Our contributions are as follows:

\smallsection{Conceptual Model}
We introduce a conceptual model that allows us to understand performance in terms of three dimensions: metadata representations, algorithms associated with \lsts and their operations, and engine characteristics (how they carry out certain operations and how they use the underlying \lsts) ~(\S\ref{sec:otfs:overview}). 
Our model provides insights into the factors that influence query performance in \lsts and highlights the surprising amount of commonality across different table formats. We hope this can help the community to develop a more unified approach to building, using and benchmarking these emerging table implementations.

\smallsection{\sys: A Benchmarking Framework for \lsts}
We design a benchmarking framework, \sys, that focuses on key characteristics of \lsts and measures fundamental metrics for a thorough understanding of their relative performance in different scenarios~(\S\ref{sec:benchmark}). Building upon TPC-DS, our benchmark proposes new extensions, relevant to \lsts, and offers the ability to define {\em packages} of workload patterns, inspired by real-world analytical workloads. We include a baseline package for completeness.

\smallsection{\sys Implementation}
We implement the framework, tying our novel benchmark and proposed metrics together~(\S\ref{sec:framework}). 
It automates the process of running the workloads and collects the required telemetry from the engine and various cloud services to compute the metrics necessary for evaluation. 
\sys is available as an open-source contribution\footnote{\url{https://github.com/microsoft/lst-bench/}.}.

\smallsection{Evaluation}
We use \sys to evaluate the performance, efficiency, and stability of out-of-the-box Delta Lake, Apache Iceberg, and Apache Hudi. Through our analysis of the results~(\S\ref{sec:evaluation}), we provide insights into their strengths and weaknesses.
Additionally, we conduct experiments using two widely adopted data processing engines, Apache Spark and Trino, and showcase their significant impact on the overall performance and efficiency of \lsts. 

\vspace{0.5em}\noindent
Our primary focus in this work is to develop a fair, comprehensive, and consistent framework for evaluating \lsts. 
While we provide insights into the current state of \lsts, it is important to note that different results could accrue for several reasons in practice: ($i$)~using a different base workload or benchmark package, ($ii$)~impact of overall system, including aspects such as security and access control, or ($iii$)~impact of engine and its level of integration with the \lst{}. 
We have designed \sys as a modular, easy-to-extend framework and welcome contributions to our open-source codebase to enhance the framework and methodology proposed in this paper.

\section{A Conceptual Model for \lsts}
\label{sec:otfs:overview}

Delta Lake, Hudi, and Iceberg are rapidly evolving independently while sharing core characteristics, such as versioned table data stored as immutable files, a metadata layer describing the data, various controls and optimizations, and integration with popular engines. 
In this section, we propose a conceptual model that utilizes these common characteristics and represents an \lst configuration along a three-dimensional space. 
Specifically, the first dimension pertains to the metadata layer representation introduced by each \lst (\S\ref{sec:model:metadata}), 
the second dimension encompasses algorithmic components and \lst parameters for controls and optimizations (\S\ref{sec:model:algorithms}), 
and the third dimension focuses on the behavior of the data processing engines (\S\ref{sec:model:engines}). 
The performance and efficiency of query processing on \lsts are determined by the combination of these dimensions. 
In \S\ref{sec:evaluation}, we rely on this model to analyze and compare the observations, similarities, and differences among the various \lsts tested.

\begin{figure*}[t]
     \centering
     \begin{subfigure}[b]{0.27\textwidth}
         \centering
         \includegraphics[width=\textwidth]{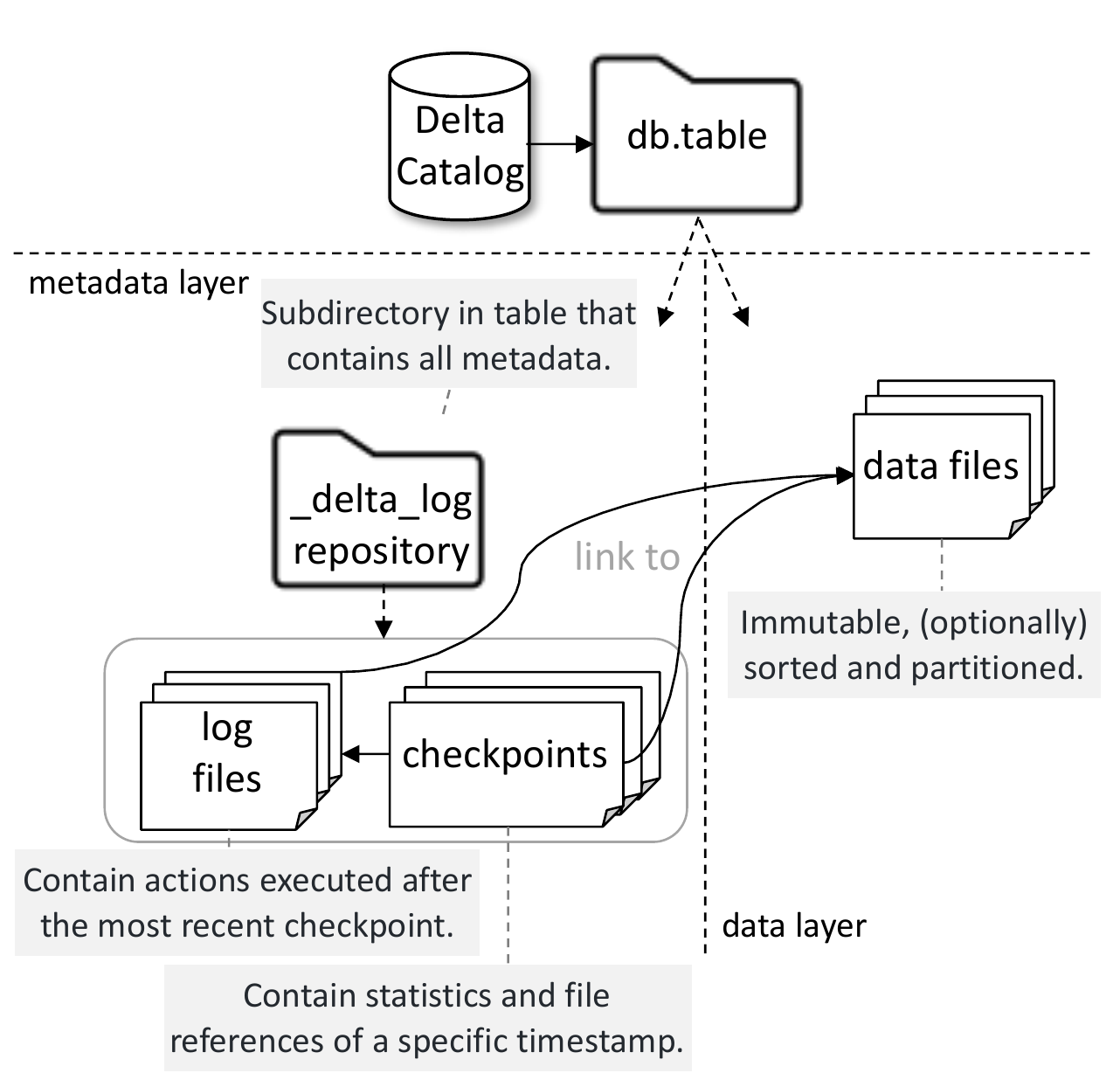}
         \caption{Delta}
         \label{fig:layout_delta}
     \end{subfigure}
     \rulesep
     \begin{subfigure}[b]{0.31\textwidth}
         \centering
         \includegraphics[width=\textwidth]{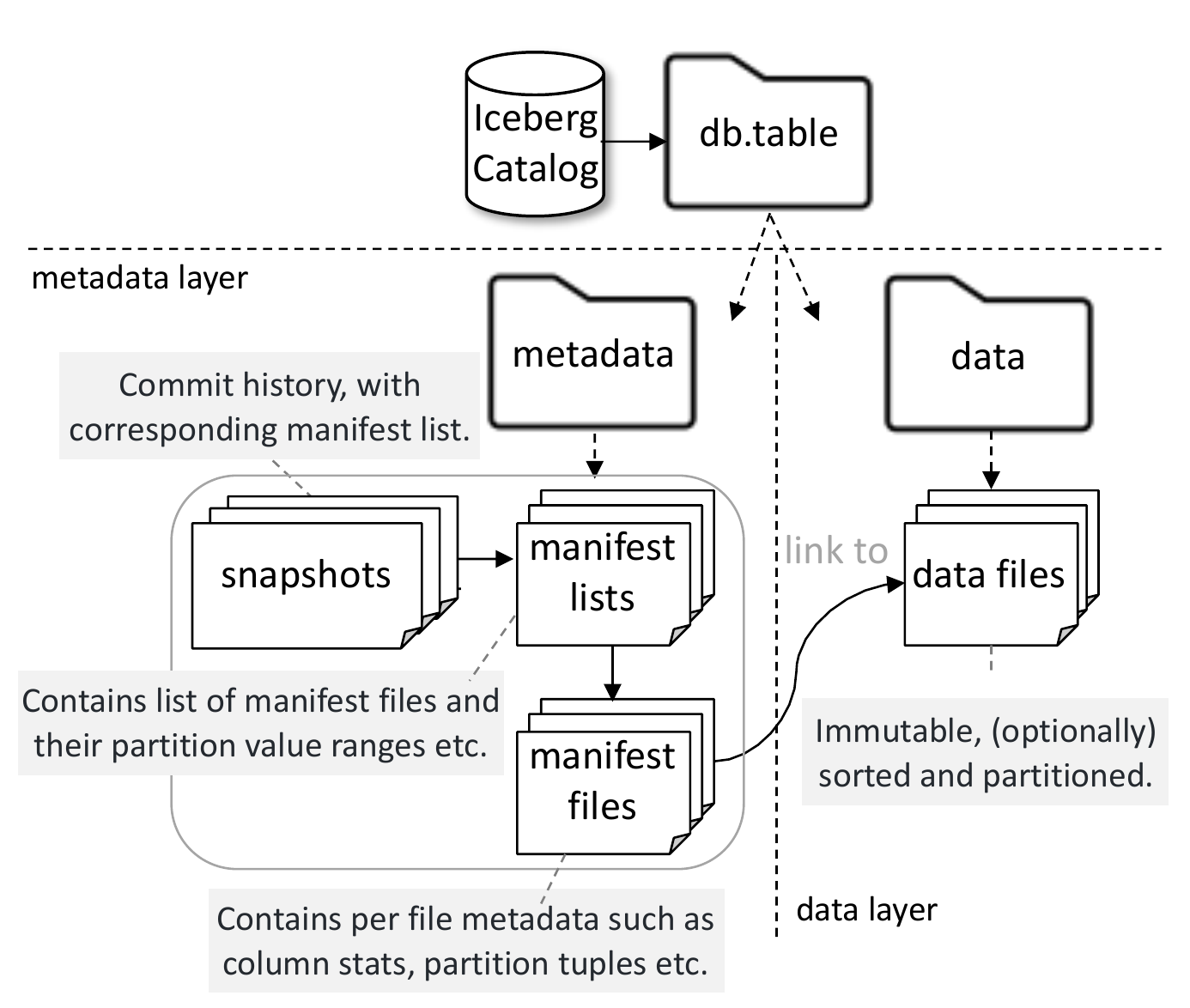}
         \caption{Iceberg}
         \label{fig:layout_iceberg}
     \end{subfigure}
     \rulesep
     \begin{subfigure}[b]{0.39\textwidth}
         \centering
         \includegraphics[width=\textwidth]{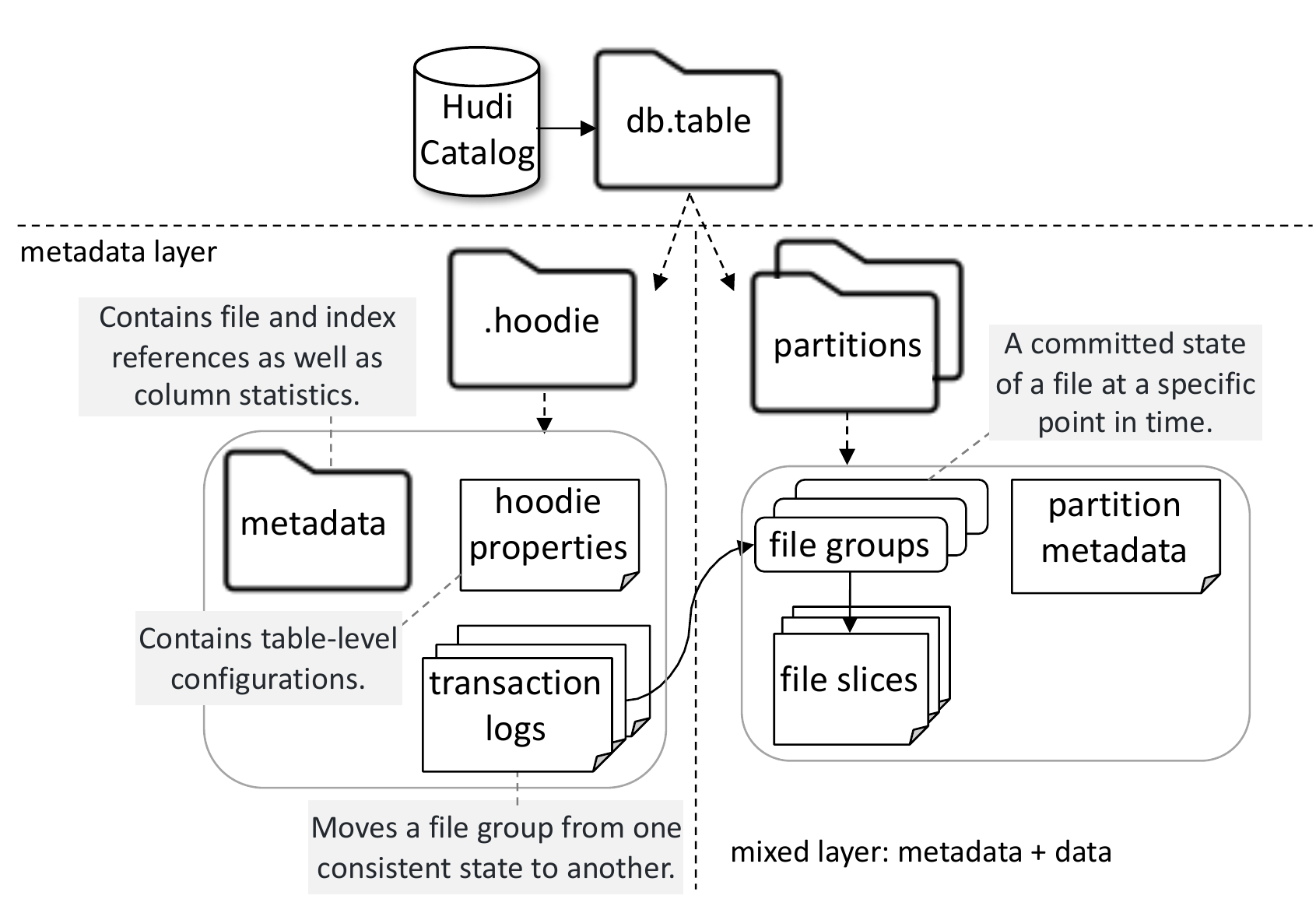}
         \caption{Hudi}
         \label{fig:layout_hudi}
     \end{subfigure}
    \caption{File layouts for \lsts under study.}
    \label{fig:file_layouts}
\end{figure*}

\subsection{Metadata}
\label{sec:model:metadata}

The metadata layer plays a crucial role in determining how engines interact with the format and serves as one of the key distinguishing components of \lsts\footnote{Notably, translating between different metadata representations enables representing one \lst as another without costly data rewriting~\cite{hudi-onetable, delta-uniform}.}. 
\lsts store metadata within the corresponding file structure of a table. 
Specifically, these formats maintain a commit log of operations performed on each table, such as adding or removing data files, or modifying the schema.
How the (meta)data files are modified may change depending on the \lst{}, and the way the metadata is laid out in the storage system is a key factor that affects the performance and features of different \lsts. 
A summary of the various metadata layouts is included in \Cref{fig:file_layouts}.

For each commit, Delta stores a log file identified by a monotonically increasing ID. This file includes an array of actions that were applied to the previous version of the table, along with statistics such as the minimum and maximum values for each column. 
The metadata subdirectory also includes \emph{checkpoint files} that store non-redundant actions. These files are generated every 10~transactions by default and are referenced in the table metadata for quick access to the last checkpoint. 
In contrast, Iceberg takes a different approach by organizing files hierarchically to represent the table's state. The top-level structure consists of a \emph{metadata file} that is replaced atomically whenever changes occur. 
This file contains references to \emph{manifest list files}, each of which represents a snapshot of the table at a specific point in time. The manifest list files, in turn, reference \emph{manifest files}, which track the data files and provide statistics about them. 
Lastly, Hudi creates a \emph{timeline} by storing the actions performed on the table as files identified by their start commit time. 
It also uses a nested metadata table~\cite{hudi-table-metadata}, 
which is a Hudi table itself, to store physical file paths and indexed files that belong to the table, thereby enabling efficient file pruning.

\subsection{Algorithms}
\label{sec:model:algorithms}

Implementations of \lsts incorporate various algorithms and configurations that give rise to different behaviors. 
These algorithms and configurations encompass protocols for engine interaction, impacting concurrency and isolation guarantees, as well as configuration parameters that influence higher-level aspects like the data layout within the table, with significant implications for query performance. 
Additionally, \lsts introduce diverse algorithms for bookkeeping and cleanup operations. 
Importantly, all these aspects are accompanied by configuration parameters that allow for fine-tuning the behavior of a particular \lst implementation.

\subsubsection{Concurrency and Locking.} 
\lsts leverage multi-version concurrency control (MVCC)~\cite{hekaton,hyper,postgres} to allow for concurrent transactions to access the same data in the storage system without interfering with each other. 
This is achieved by creating multiple versions or \emph{snapshots} of the same logical data.

All three \lsts allow transactions to be executed within the context of a single table by implementing optimistic concurrency control~\cite{delta-occ,iceberg-occ,hudi-occ}. However, they differ in their approaches to conflict resolution and the level of isolation they provide. 
Hudi and Iceberg support \emph{snapshot isolation}, which means that even if other transactions are concurrently modifying the table, a transaction reads a consistent snapshot of the data as it existed at the start of the transaction. 
In turn, Iceberg and Delta provide a stricter isolation level by default, which guarantees that writes to the table will occur in a serial order. 

The lock management requirements, and consequently their implementation, differ across the three \lsts due to their different designs. Delta and Iceberg have minimal lock management needs and rely only on atomic \textit{put-if-absent} or \textit{rename} operations provided by the underlying object store or file system~\cite{DBLP:journals/pvldb/ArmbrustDPXZ0YM20,iceberg-spec}. Hudi has a greater reliance on locking, particularly when using the metadata table to track table files~\cite{hudi-multi-writer}.

\subsubsection{Data Layout Configuration.} 
Since \lsts operate on the assumption that data files are immutable, they require mechanisms for updating and deleting rows. 
The two supported strategies are Copy-on-Write (CoW) and Merge-on-Read (MoR). 
CoW creates a new copy of the data files for each update or delete operation, while MoR writes changes to a separate file (often referred to as \emph{delta file}) that is merged into the dataset during read operations. 
CoW is preferred for read-heavy workloads, while MoR is the preferred strategy for write-heavy workloads. 
Iceberg and Hudi currently support both CoW and MoR, while Delta only supports CoW\footnote{MoR support is an upcoming feature in Delta Lake.}.

\subsubsection{Table Maintenance Operations.} 
\lsts, similar to previous MVCC implementations~\cite{inmemory-mvcc}, provide operations that ensure the data stored in a table is optimized and efficient. 
These operations include ($i$)~\emph{compacting} data files within a table, which consolidates smaller files into larger ones and optionally sorts the data based on specific column values, reducing metadata overhead and improving query performance, and ($ii$)~\emph{vacuuming} data files within a table, which deletes expired data files after the retention period or ones that are no longer referenced by the table metadata because they are deemed useless after the compaction process. 

Engines relying on \lsts typically provide an API to perform these maintenance operations on-demand. For instance, commands to execute them are often available to users as SQL extensions~\cite{spark-delta-optimize} or stored procedures~\cite{spark-iceberg-optimize,spark-hudi-optimize}. 
\lsts have different default settings for using these maintenance operations, for example, Hudi enables compaction and vacuum out-of-the-box while Iceberg and Delta require users to specify them.
Users may execute these operations after modifying table data or schedule them to run automatically based on predetermined criteria such as the age or size of the data files. 
Additionally, some commercial platforms offer the automation of table maintenance~\cite{databricks,tabular,onehouse}.

\subsection{Engines}
\label{sec:model:engines}

Engines and their configurations have a significant impact on the performance of all \lsts. 
Clearly, the efficiency of engine internals directly affects the speed of data read and write operations to the \lst. 
However, there are additional aspects related to engine configuration that may not be immediately apparent but have substantial implications for performance.

For example, the cluster configuration, parallelism settings, and the chosen execution plan by the engine can have effects beyond the execution of individual queries. 
Concretely, these factors can impact the fragmentation at the storage level, as they can influence the number of files generated by the engine during write operations (recall our discussion comparing the behavior of Spark and Trino in \S\ref{sec:intro}). 
This, in turn, has ripple effects on subsequent operations, affecting the performance of both reads and writes, including those executed during table maintenance operations. 

\section{\sys Benchmarking Framework}
\label{sec:benchmark}

TPC-DS~\cite{tpcds-refresh} is widely used to evaluate decision support systems, covering various aspects such as data loading, query execution, sustained throughput, and data updates. 
Researchers and practitioners are also extensively using its dataset and workload to assess the efficiency of \lsts~\cite{brooklyn-data,lhbench,transparent-benchmarks,reassessing-performance}, often measuring query latency. 
In this section, we propose a new benchmark for \lsts that builds upon TPC-DS as a ``base workload'', using its data set generator as well as its query set but modifying how the benchmark execution is structured. 
Specifically, we create several workload patterns that invoke tasks of the base workload along with other \lst{} specific tasks such as compaction or time travel (\S\ref{sec:benchmarks}).
Furthermore, we propose new metrics intended to capture performance characteristics specific to \lsts (\S\ref{sec:measurements}).

\subsection{Workload Patterns}
\label{sec:benchmarks}

We divide the work of extending the base workload into two parts. 
First, we enhance the benchmark's customizability by proposing new tasks specific to \lsts, as well as making it easier to use existing ones to create custom workloads. This is discussed in detail in \S\ref{sec:benchmarks:extensions}.
Second, we utilize the proposed extensions to create a baseline \emph{package} consisting of workload patterns useful to gain insights into \lst aspects overlooked by the base workload, such as \emph{stability}, \emph{resiliency}, \emph{read/write concurrency}, and \emph{time travel}. These workload patterns are presented in \S\ref{sec:benchmarks:tfe}. 
Before discussing these extensions in detail, we introduce the model employed for workload representation and, for completeness, briefly describe the original TPC-DS workload. 

\begin{figure}[t]
\centering
\includegraphics[width=\columnwidth]{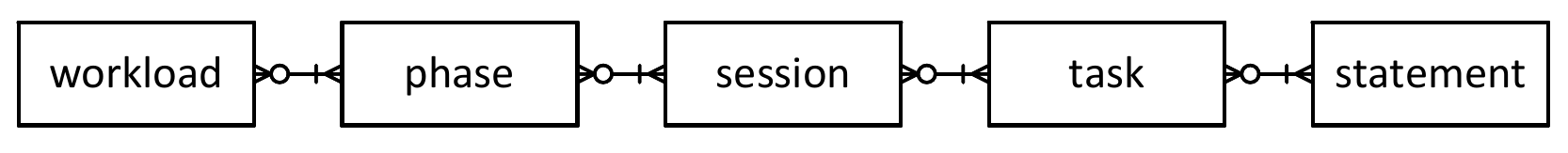}
\caption{Workload components and their relationships.}
\label{fig:diagram_workload}
\end{figure}

\begin{figure*}[t]
\centering
\includegraphics[width=\textwidth]{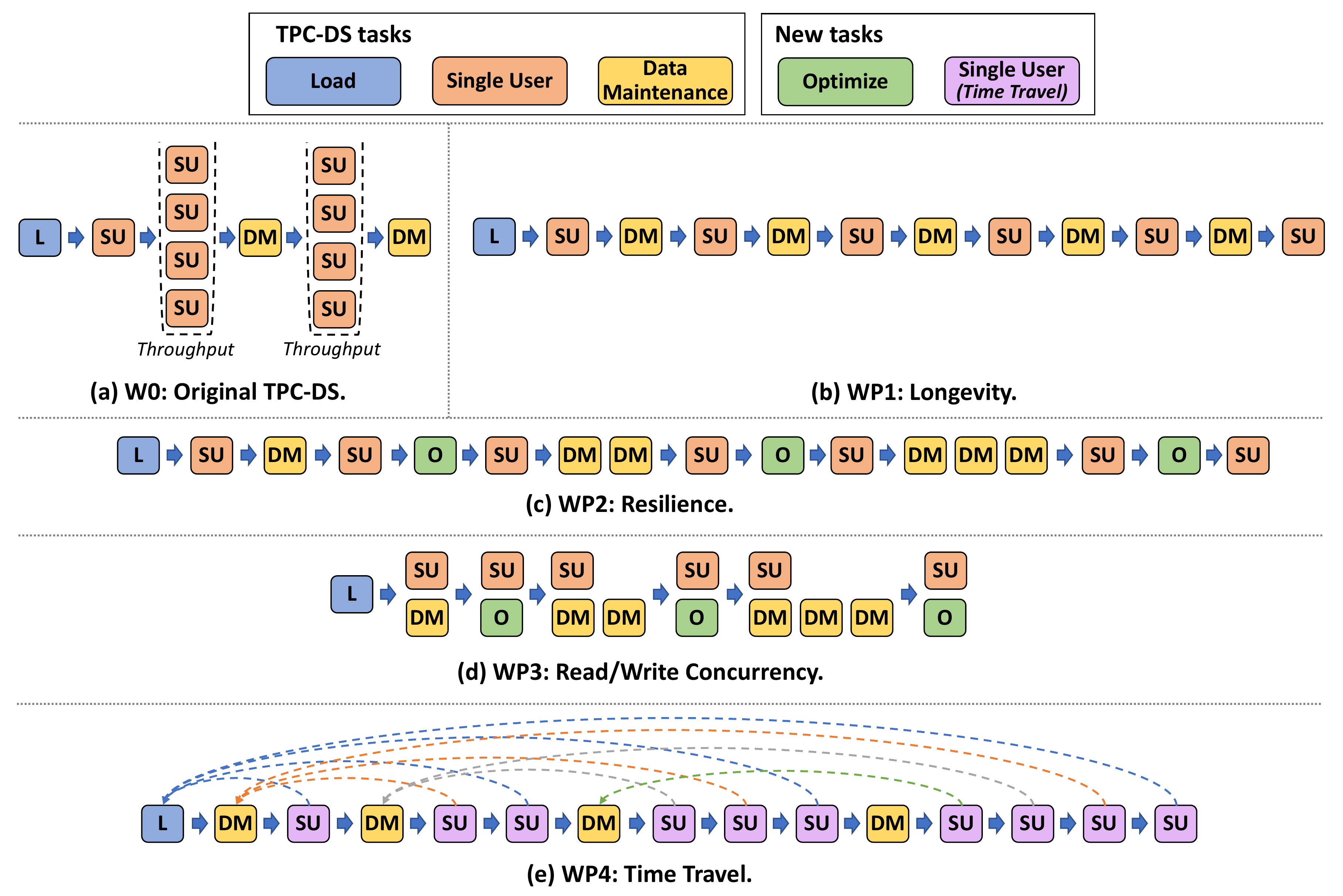}
\caption{TPC-DS and extensions to evaluate \lsts characteristics.}
\label{fig:tpcds-ext}
\end{figure*}

\smallsection{Workload Representation}
\Cref{fig:diagram_workload} depicts the components of a workload and their relationships. 
A \emph{task} is a sequence of SQL \emph{statements}, while a \emph{session} is a sequence of tasks that represents a logical unit of work or a user session. 
A \emph{phase} is a group of concurrent sessions that must be completed before the next phase can start. 
If a phase consists of a single task, we may refer to it interchangeably by its task name. 
Lastly, a \emph{workload} is a sequence of \emph{phases}. 

We choose this flexible representation to ensure its adaptability to both standard and custom workloads prevalent in practice, offering a comprehensive solution. 
Although not all scenarios require all abstractions, our approach was driven by diverse and extensive user feedback, including ($i$)~facilitating the mapping of existing workloads such as TPC-DS, ($ii$)~aligning with the concept of sessions in JDBC, and ($iii$)~ease of reusability. 
For instance, a session initiates a fresh connection to the engine, whereas a task simply groups a collection of SQL statements. 
Consequently, a workload designer can either execute multiple tasks in one session or initiate numerous concurrent sessions, each handling a subset of tasks.

\smallsection{\baseworkload. \emph{Original TPC-DS}}
According to these definitions, the TPC-DS benchmark executes multiple phases as shown in Figure~\ref{fig:tpcds-ext}a. 
These phases include ($i$)~a \load phase where data is loaded into the tables used in the experiment, ($ii$)~a \singleuser phase which runs a series of complex queries to determine the upper limit of the engine's performance, ($iii$)~\throughput phases involve running multiple sessions in parallel, each executing a \singleuser task with a different permutation of the query set, to assess the engine's ability to handle multiple users and queries simultaneously, and ($iv$)~\datamaintenance phases that are executed to test the engine's ability to handle data inserts and deletes.

\subsubsection{Workload Composability.}
\label{sec:benchmarks:extensions}

Next we describe our extensions to enhance TPC-DS customizability.
Note that although we will propose specific workload patterns in \S\ref{sec:benchmarks:tfe}, our extensions offer flexibility for developing new patterns that can highlight characteristics that may have been overlooked in previous evaluations.

\smallsection{Configurable Sequence of Phases} As mentioned previously, the TPC-DS standard defines a strict sequence of phases that must be executed in a specific order. To evaluate specific aspects of \lsts, for example their longevity, we require a more flexible approach to the order of phases that is not captured by the standard sequence. For this reason, the sequence of phases that the benchmark executes should be configurable.

\smallsection{Ability to Run Multiple Tasks Concurrently within a Phase} TPC-DS sequences are linear, meaning that different tasks never overlap with each other (even though in the \emph{throughput} phase multiple \singleuser tasks run in parallel). However, as other works have previously reported~\cite{armbrustmaking}, a common use case for \lsts is querying the data while background operations, such as the incremental maintenance of downstream tables or materializations, are concurrently executing. Therefore, we want to be able to evaluate \lsts while running multiple, possibly different, tasks concurrently.

\smallsection{\phase{\textbf{Optimize}} Task} Given that table maintenance operations are frequently executed concurrently with other queries, it is critical to include them in the evaluation of \lsts to determine whether they ($i$)~can restore the \lst to its initial non-degraded performance state, and ($ii$)~impact the performance of other queries running concurrently. 
To address this, we introduce a new \optimize task that involves running \emph{compaction} on \lsts. 
While \lsts offer various \emph{compaction} strategies to optimize file layout and size, such as bin-packing or sorting, we opt to use the default strategy for each \lst in our task definition. 
However, note that selecting an alternative strategy would be as straightforward as modifying the SQL associated with the task definition.

\smallsection{\phase{\textbf{Time Travel}} Task} In \S\ref{sec:intro:prob}, we mentioned that a benchmark should evaluate new features provided by \lsts, such as \emph{time travel}, which allows querying of historical versions of a table based on timestamp or version. Therefore, we introduce a \timetravel task that executes the same queries as a \singleuser task but as of a given point in time. 

\subsubsection{Baseline Package for Evaluation of \lsts.}
\label{sec:benchmarks:tfe}

Based on these extensions, we propose a package consisting of four workload patterns to gain insights that cannot be obtained by executing the original TPC-DS workload. To design our experiments, we carefully selected various parameter values, including experiment length, based on empirical observations from customer workloads.

\smallsection{\workload{1}. \emph{Longevity}}\label{sec:benchmark_w0} This workload evaluates the performance, cost, and IO stability of \lsts over time. The proposed sequence is shown in Figure~\ref{fig:tpcds-ext}b. The experiment involves six \singleuser phases, each followed by a \datamaintenance phase to add new metadata and data to the table. Repeating this process multiple times allows us to observe how the \lst behaves over time and identify significant trends, if there are any.

\smallsection{\workload{2}. \emph{Resilience}} This workload, shown in Figure~\ref{fig:tpcds-ext}c, evaluates the impact of table maintenance operations such as compaction on degradation over time. Each \optimize phase is executed subsequent to an increase in the number of write statements executed on the source tables, thus measuring the performance of \optimize operations as the ratio of refreshed data in a table increases.

\smallsection{\workload{3}. \emph{Read/Write Concurrency}} This workload evaluates the impact of the concurrent execution of read and write statements. As shown in Figure~\ref{fig:tpcds-ext}d, we run \singleuser phases concurrently with the \datamaintenance and \optimize phase respectively to simulate this scenario. Note that by leveraging the separation of storage and compute and the on-demand availability of cloud computing resources, our framework has the ability to run concurrent operations on separate compute clusters, which allows us to evaluate the storage layer's impact without the complication of interleaving these operations at the compute layer.

\smallsection{\workload{4}. \emph{Time Travel}} 
\lsts introduce \emph{time travel}, which enables querying data at specific points in time by leveraging SQL extensions to specify the desired version of the table~\cite{delta-time-travel,iceberg-time-travel,hudi-time-travel}. 
The workload shown in Figure~\ref{fig:tpcds-ext}e evaluates this new feature. 
We execute multiple \datamaintenance phases on the original data, followed by the same number of \timetravel phases, each executed on a version of the table produced by a previous \load or \datamaintenance phase.

\myparagraph{Discussion.}
We could enhance our proposal through additional extensions, such as new tasks to cover dimensions that are still overlooked in the current proposal, e.g., \emph{vacuum} operations, schema evolution~\cite{delta-schema-evolution,iceberg-schema-evolution,hudi-schema-evolution}, partition evolution~\cite{iceberg-partition-evolution}, or deep and shallow table cloning~\cite{delta-clone}. 
Moreover, we could introduce tasks containing SQL statements that modify engine behaviors at the session level (e.g., Spark's \emph{set} statements~\cite{spark-set-statements}), facilitating the evaluation of important engine features and configurations, including parallelism settings. 
Lastly, incorporating workload patterns consisting of a more concurrent, diverse, and intricate mix of tasks would further assist in evaluating \lsts. 
For instance, this could help to evaluate scenarios characterized by increased concurrency, which is common when dealing with \lsts, and diverse conflict resolution and isolation level configurations. 
It could also help to assess whether background operations like file consolidation and clustering, which typically run automatically and continuously in fully managed platforms without explicit invocation during a workload's execution, are indeed triggered and yielding their intended effects. 
Implementing these extensions into our framework is straightforward, and we plan to explore it in the future.

\subsection{Metrics}
\label{sec:measurements}

This section explores metrics for a comprehensive and fair evaluation of \lsts unique features. 
We first discuss traditional metric categories applicable to \lsts, such as performance, and storage and compute efficiency (\S\ref{sec:measurements:primitive}). 
We then introduce a stability metric that builds upon the aforementioned metrics to reveal crucial degradation characteristics of cloud data warehouses (\S\ref{sec:measurements:stability}). 
Our multi-metric approach draws from prior research~\cite{ycsb,tpc-benchmarks} and our own observations, and importantly, it can be easily extended to cover unexplored dimensions.

\subsubsection{Traditional Metrics.}
\label{sec:measurements:primitive}

In a cloud environment, several categories of metrics are important to consider when evaluating \lsts.
First, data warehouse performance is traditionally evaluated using two measurements, latency, i.e.,~measuring the round-trip of queries, and throughput, i.e.~, measuring the capacity of the system.
Second, the interaction of a \lst with the storage layer is an important aspect as \lsts specifically rely on cloud (object) storage.
Unlike local disks, cloud IOPS are charged on a pay-as-you-go basis.
This means that managing storage utilization is not the only factor to consider, but also total API operations and data transfers, and peak rates.
Finally, a compute efficient \lst achieves high performance while using a small amount of resources which is especially crucial on shared clusters.
Key metrics include CPU utilization, memory utilization, and disk utilization, which measure the amount of CPU, memory, and local disk used for processing the workload, respectively.

Note that some of these (types of) metrics are captured as part of the TPC-DS standard.
For example, QphDS is a throughput metric while load time, system availability and price per QphDS are additional metrics that contribute to a more comprehensive understanding of the evaluated systems.

\subsubsection{Stability.}
\label{sec:measurements:stability}

\lsts are designed to receive continuous trickle updates, which, over time, can result in accumulation of delta files in the object store. 
Intuitively, the oftentimes smaller delta files degrade the system's efficiency as it causes the compute to consume additional resources to successfully execute a workload. The extent of the degradation depends on several factors, including the number of new files and data layout in the files. A well designed system is less susceptible to the degradation as more data updates are performed, or may have features to auto-mitigate adverse side effects of updates. To measure degradation, we introduce a new metric category, \emph{stability}, which examines a system's ability to sustain its performance and efficiency (e.g., latency) consistently and exhibit minimal degradation. The process of calculating degradation involves dividing a workload's timeline into different phases, as described in \S\ref{sec:benchmarks}, and then comparing the performance and efficiency measurements taken during each phase of the same type. For example, \singleuser phase SU-$i$ would only be used for the computation of stability pertaining to \singleuser phase performance. We formally define the degradation rate below.
\begin{equation} \label{eq:stability}
S_{DR} = \frac{1}{n}{\sum_{i=1}^n \frac{M_i - M_{i-1}}{M_{i-1}}}
\end{equation}
where
\begin{itemize}
    \item \(M_i\) is metric value of the \(i^{th}\) iteration of a workload phase,
    \item \(n\) is the number of iteration of the phase, and
    \item \(S_{DR}\) is the degradation rate. 
\end{itemize}
Intuitively, \(S_{DR}\) is the rate at which a metric is growing or shrinking, due to the cumulative effects of changes in the underlying system's state. It provides information about how quickly a system degrades. A \emph{stable} \lst exhibits low \(S_{DR}\). 
Note that $M$ can be selected from the metrics introduced in \S\ref{sec:measurements:primitive}; for metrics where higher values indicate better performance (e.g., throughput), the same function can be used by replacing $M$ with its reciprocal $1/M$.

\begin{figure}
    \centering
    \includegraphics[width=\columnwidth]{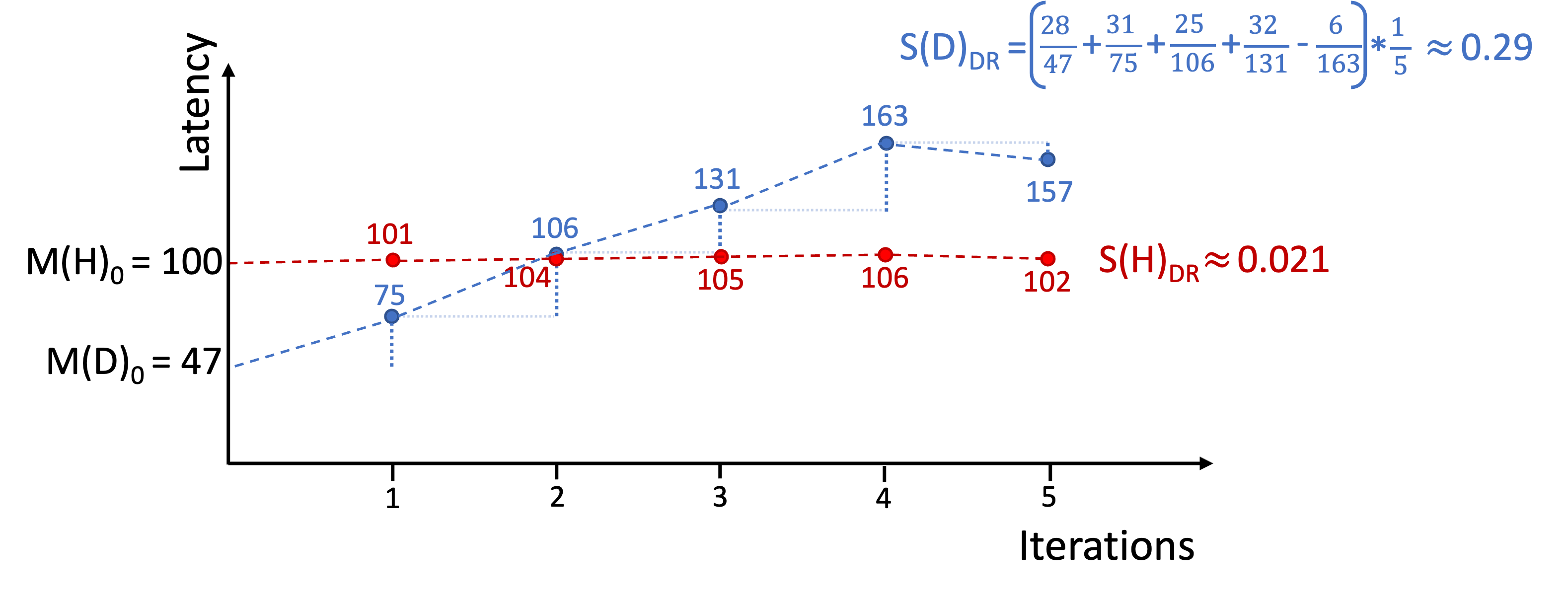}
    \vspace{-1em}\caption{Example for \emph{stability} computation.}
    \label{fig:meas:stability}
\end{figure}

\begin{example}
\Cref{fig:meas:stability} shows $S_{DR}$ evaluation of a system $D$, $S(D)_{DR}$, and a system $H$ $S(H)_{DR}$ base on latency measurements over time. For $D$, we calculate the degradation rate as $(\frac{28}{47}+\frac{31}{75}+\frac{25}{106}+\frac{32}{131}-\frac{6}{163}) * \frac{1}{5} \approx 0.29$, while $S(H)_{DR} \approx 0.021$. This indicates that system $D$ is less stable than system $H$, and, without any mitigation actions, it will under perform over time. 
\end{example}

\myparagraph{Discussion.}
Other reasonable metrics can be incorporated in this scheme. One example is relative standard deviation, which remains impartial to the sequence in which the measurements are presented. We opted for the current metric because it forms an essential building block for making predictions and taking appropriate actions. Over and above the particular metric used to capture stability, we want to emphasize the importance of stability as a new characteristic to measure and optimize for.

\section{\sys Benchmarking Tool}
\label{sec:framework}

This section presents the implementation details of \sys, a tool designed to benchmark and compare \lsts in the cloud, building on the ideas discussed in \S\ref{sec:benchmark}. 
Similar to existing benchmarking systems like BenchBase~\cite{oltpbench,benchbase} and DIAMetrics~\cite{diametrics}, \sys includes an application written in Java that executes SQL workloads against a database management system using JDBC (\S\ref{sec:framework:app}).
Moreover, \sys features a processing module written in Python that consolidates experimental results and calculates metrics to provide insights into \lsts and cloud data warehouses (\S\ref{sec:framework:tiers}).

\subsection{Client Application}
\label{sec:framework:app}

\begin{figure}
    \centering
    \includegraphics[width=\columnwidth]{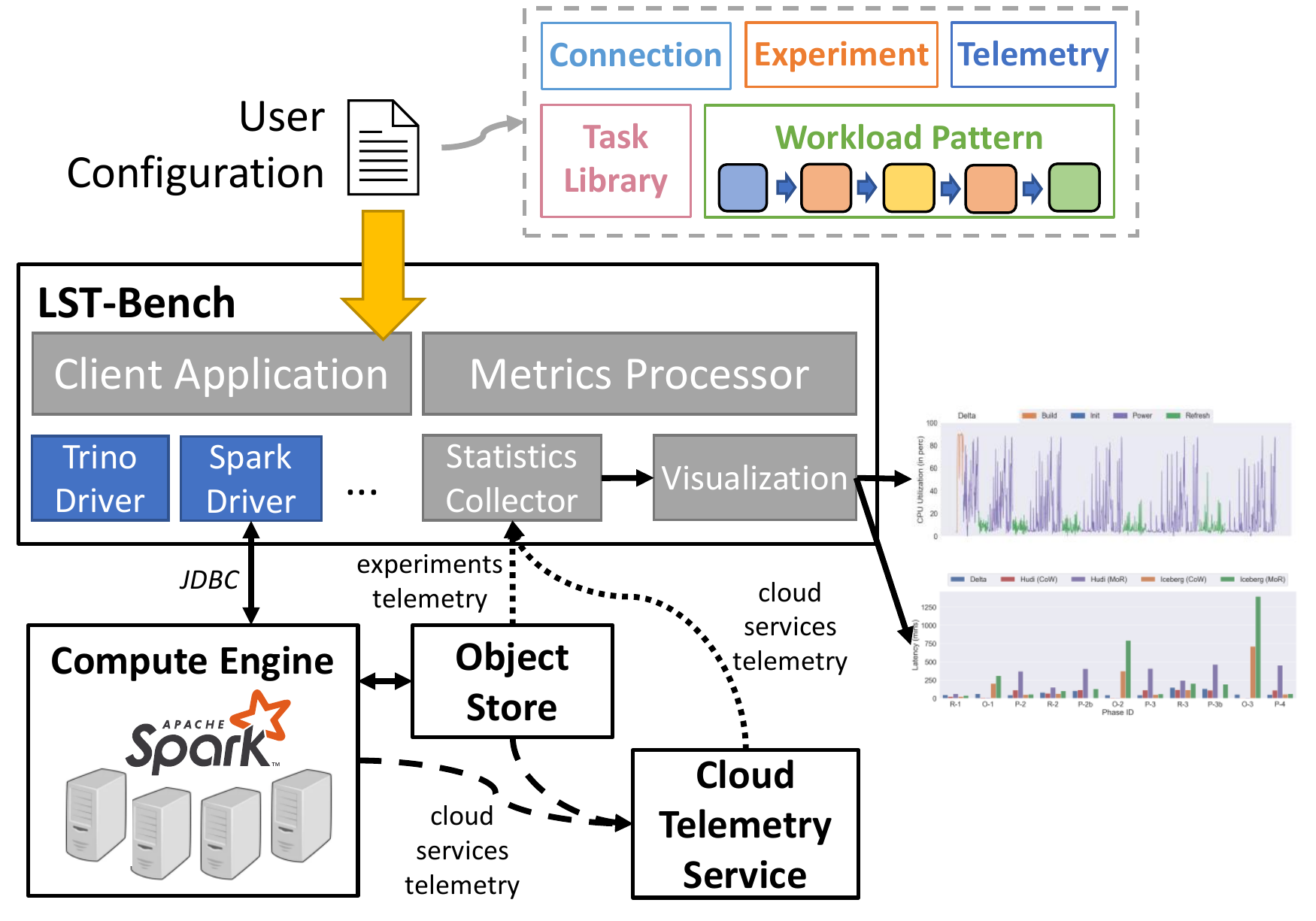}
    \caption{\sys components and execution model.}
    \label{fig:frame:framework}
\end{figure}

\sys's client is a flexible and modular benchmarking application that enables users to easily combine various configurations, \lsts, and workloads. 
As depicted in \Cref{fig:frame:framework}, it follows a \emph{configuration-driven approach} that allows users to define ($i$)~clusters connection details, ($ii$)~specific options for the experiment, including System-Under-Test (SUT) and \lst-Under-Test (LUT), ($iii$)~telemetry collection configuration, and ($iv$)~the workload to execute. 

\smallsection{Experiment Definition}
The configuration APIs enable users to define the workload for an experiment as a series of phases, with each phase identified by a unique, user-defined name. 
\sys provides a library containing the TPC-DS and new tasks described in \S\ref{sec:benchmarks}.
Each task consists of a sequence of SQL statements stored in files in a folder hierarchy; there are different files for the various SQL dialects supported. 
In addition, certain tasks, such as \optimize, have separate files for each \lst under each engine, as \lsts have distinct syntax for executing their table maintenance operations.
\sys selects the appropriate files for an experiment based on the workload configuration provided by the user.

\smallsection{Customizability and Extensibility}
The SQL files can contain variables that \sys replaces before executing a particular task. 
Users can use this mechanism to pass configuration values to a task, such as the catalog name, database name, or desired table location in the object store. 
Additionally, \sys utilizes this feature internally to inject necessary information to run certain tasks, such as the timestamp value for \timetravel. 

Incorporating new tasks into the \sys library is straightforward. 
Users can add the files with the SQL statements that need to be executed and include the new task in the \sys library. 
Once done, the new task can be referenced from the workload definition files.

\smallsection{Experiment Execution}
It is important to note that \sys does not automatically deploy and configure the compute cluster; we are exploring this extension as future work. 
Instead, it is the user's responsibility to deploy the engine with the corresponding \lst libraries. 
When conducting an experiment, \sys utilizes JDBC and engine-provided drivers to connect to the SUT. 
\sys creates a \emph{JDBC connection pool} of a size equal to the maximum number of concurrent sessions required by any phase in the experiment. 
The workload configuration contains the phases that need to be executed sequentially during the experiment. 
Moreover, the default library includes tasks definitions that create external tables in the engine catalog. 
These tables point to object store directories containing TPC-DS data generated by the benchmark tool at the specified scale factors. 
These tasks are typically invoked at the start of an experiment and the external tables they create are currently used in \load and \datamaintenance tasks.

\subsection{Metrics Processor}
\label{sec:framework:tiers}

The metrics processor relies on telemetry collected by \sys as well as external, cluster-level telemetry that the user enables up through cloud services.
Specifically, \sys collects and stores a detailed breakdown of the start and end times of each experiment, phase, task, and statement, as well as other configuration values during the execution of a given workload.
This telemetry allows us to reason about the latency and throughput of the evaluated \lsts.
In addition, we rely on time series data that captures resource utilization, storage API calls, or network I/O volume gathered by cloud service providers which is accessible via dedicated APIs\footnote{\sys leverages standard auditing telemetry, making its insights-gathering approach broadly applicable. However, in scenarios where access to telemetry from cloud services is restricted, such as when the platform is offered through a vendor, certain insights might be unavailable.}.
The metrics processor package provides generalized drivers for extracting external telemetry as well as specific implementations of those cloud setups that we use for our evaluation.
It also provides notebooks and templates that allow users to plot the same (types of) figures capturing both internal and external telemetry that we will discuss next in our evaluation.

\section{Evaluation}
\label{sec:evaluation}

In this section, we present benchmarking results for the workloads described in~\S\ref{sec:benchmarks} using \sys with Delta Lake, Apache Iceberg, and Apache Hudi, running on Apache Spark~\cite{apache-spark} and Trino~\cite{trino}. {\em We used default parameter settings (e.g., isolation level) and did not perform any special tuning for the evaluated \lsts. Tuning for optimal performance is beyond the scope of this paper, whose focus is on how to benchmark performance over \lsts.}
We chose Spark and Trino as the compute engines for the evaluation because they are widely adopted, open-source, and offer the most mature integration with the \lsts examined in this study, based on our own experience. We note that our benchmarking framework can readily be used with other engines, and the extent to which those engines are integrated with different \lsts will materially impact the performance across \lsts.  In brief, our results show:
\begin{itemize}[leftmargin=*]
\item The accumulation of data files significantly degrades \lst's performance, up to $6.8x$ in our study, unless maintenance is performed to mitigate its impact.
\item DML operations on Spark can result in a significantly higher number of delta files than Trino, causing up to a 2.4X degradation in performance in our tests. Additionally, the baseline query workloads run at nearly double the speed on Trino for both Delta and Iceberg tables.
\item CoW and MoR modes have significant trade-offs regarding their read/write interaction with the storage layer. For example, Hudi and Iceberg MoR on Spark lead to high I/O volume and calls, respectively, resulting in higher read query latency than CoW.
\item Table maintenance has a big impact on Delta and Iceberg performance stability, whereas Hudi maintains stable performance without periodic maintenance by doing more upfront work.
\item Tuning \lsts involves trade-offs depending on user goals. For example, Iceberg's default file group-by-group compaction reduces disruption on read queries running on the same cluster, but significantly increases compaction time.
\item Concurrent read/write sessions have non-trivial impact on query performance. Combining maintenance operations with read queries on the same cluster can improve resource utilization without affecting read latency. Running multiple compute clusters concurrently can reduce execution time by leveraging compute and storage decoupling in cloud engines.
\end{itemize}

We want to emphasize that the results we report are specific to the versions and configurations that we tested, and their performance can be subject to change and improvement due to further tuning and future developments. 
Our main objective in sharing these findings is to demonstrate \sys's ability to quantify noteworthy trade-offs across combinations of engines and \lsts.

Our results highlight an important point--each \lst offers opportunities to tune performance by making careful choices, e.g., when to do maintenance. In general, these choices depend upon the target workload.

\myparagraph{Hardware and Software Setup.}
Our experiments were conducted on clusters running Spark 3.3.1 and Trino 420. 
Each cluster comprised 1 head and 16 worker nodes. 
Furthermore, for the evaluation of concurrency (\S\ref{sec:eval:concurrency}), additional clusters consisting of 1 head and 7 worker nodes were used. 
All clusters were provisioned by Azure VMSS~\cite{azure-vmss} %
and their nodes were Azure Standard E8as v5 instances with AMD EPYC\texttrademark~7763 CPU @ 2.45GHz (8 virtual cores) and 64GB RAM. 
For Spark, we used Delta Lake v2.2.0, Apache Iceberg v1.1.0, and Apache Hudi v0.12.2. 
In contrast, in Trino, the \lst implementation is integrated within the engine; currently, Trino only supports read and write operations for Delta~(CoW) and Iceberg~(MoR). 
The data sets for evaluation were stored in Azure Data Lake Storage Gen2 (ADLS)~\cite{azure-adls}. 
We leveraged Azure Monitor~\cite{azure-monitor} to collect telemetry from the compute cluster and data storage, and relied on Logs Analytics~\cite{azure-monitor-log-analytics} to execute queries against the collected data.

\myparagraph{Experimental Setup.}
To generate data at different scale factors (SF100, SF1000), we used the \textit{dbgen} tool in the TPC-DS benchmark~\cite{tpcds-refresh} and stored the generated data in ADLS. 
Data streams for \datamaintenance were also generated using the same tool and stored in ADLS. 
Our \singleuser task is a permutation of the 99~queries in the benchmark. 
For our evaluation, we use the workload patterns described in~\S\ref{sec:benchmarks}, running \workload{1} on Spark and Trino, and the remaining patterns solely on Spark; running them on Trino is left for future work.
We discuss the results of our evaluation next.

\begin{figure*}
    \centering
    \begin{subfigure}[b]{0.5\textwidth}
        \centering
        \includegraphics[width=\textwidth]{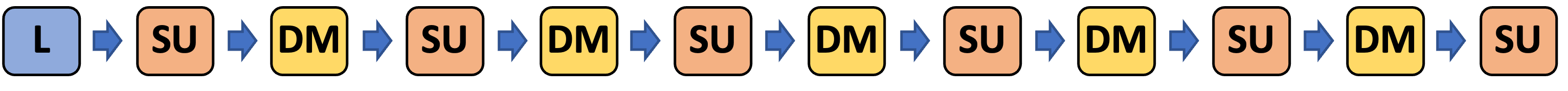}
    \end{subfigure}
    \hfill
    \begin{subfigure}[b]{0.9\textwidth}
        \centering
        \includegraphics[width=\textwidth]{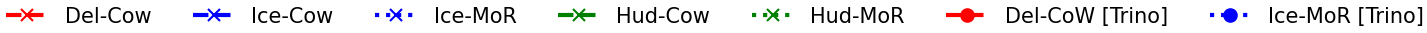}
    \end{subfigure}
    \hfill
    \begin{subfigure}[b]{0.30\textwidth}
        \centering
        \includegraphics[width=\textwidth]{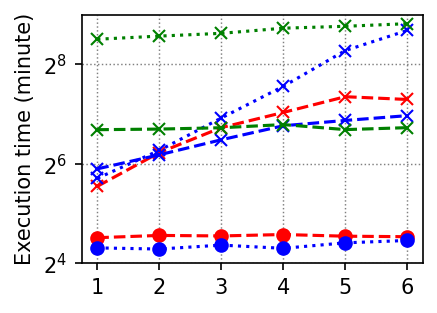}
        \caption{Performance of \workload{1} SU phases.}
        \label{fig:w1_perf_1k_su}
    \end{subfigure}
    \hfill
    \begin{subfigure}[b]{0.30\textwidth}
        \centering
        \includegraphics[width=\textwidth]{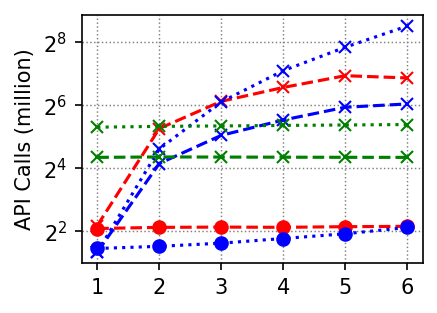}
        \caption{Total ADLS API calls by SU phases.}
        \label{fig:w1_storage_api_su_1k}
    \end{subfigure}
    \hfill
    \begin{subfigure}[b]{0.30\textwidth}
        \centering
        \includegraphics[width=\textwidth]{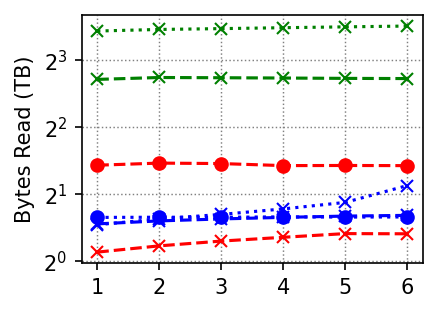}
        \caption{Bytes read from ADLS during SU runs.}
        \label{fig:w1_storage_io_su_1k}
    \end{subfigure}
    \hfill
    \caption{Evaluation of runtimes, network round trips, and storage utilization of \singleuser phases for 7 \lsts setups using \workload{1} ($SF1000$). The results highlight how increase in network round trips affects phase execution times.}
    \label{fig:w1_1k_su_degradation}
\end{figure*}

\begin{figure}
    \centering
    \begin{subfigure}[b]{0.47\columnwidth}
        \centering
        \includegraphics[width=\textwidth]{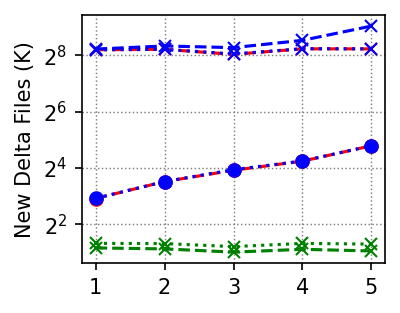}
        \caption{Files created by DM phases.}
        \label{fig:w1_new_files_dm_1k}
    \end{subfigure}
    \quad
    \begin{subfigure}[b]{0.47\columnwidth}
        \centering
        \includegraphics[width=\textwidth]{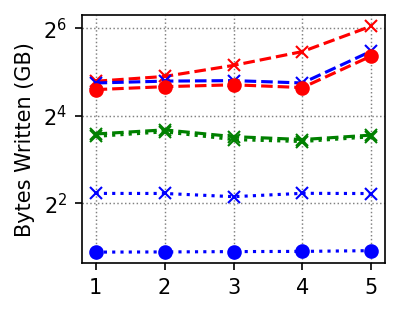}
        \caption{Total size of new files.}
        \label{fig:w1_storage_io_dm_1k}
    \end{subfigure}
    \caption{Evaluation of storage usage of \workload{1} \datamaintenance phases ($SF1000$).}
    \label{fig:w1_1k_dm_degradation}
\end{figure}

\subsection{Longevity}
\label{sec:6.1_w1}
The aim of the \emph{longevity workload}~(\workload{1}, \S\ref{sec:benchmark_w0}), consisting of six \singleuser and five interleaved \datamaintenance phases, is to evaluate an \lst's ability to maximize performance, efficiency, and stability metrics in scenarios that involve frequent data updates. Our findings demonstrate that by executing the workload using \sys, we can detect crucial configurations across all dimensions of the \lst model that significantly influence these objectives.

We ran \workload{1} against seven variations of \lsts derived from representative selections across the three dimensions of the \lsts (\S\ref{sec:otfs:overview}); Spark and Trino engines, CoW and MoR data algorithms, and Delta, Iceberg, and Hudi as metadata layouts.
Two scale factors, 100GB and 1TB, were used against each variation to account for the impact of scaling. We recorded latency of each phase within a run, along with storage and compute tier metrics~(\S\ref{sec:measurements}). 

($i$)~\emph{Compared to Trino, Spark-related variations exhibit low stability.}
\Cref{fig:w1_perf_1k_su} illustrates the execution time of all \singleuser phases in \workload{1}. 
We observe that in Spark with Delta and Iceberg, the execution time of a \singleuser phase is always higher compared to its previous run. 
This behavior is due to \emph{Spark's default distribution-mode} parameter, which results in distribution of writes in a table partition up to $200$ times~\cite{spark-distribution-mode} , creating hundreds of thousands of new delta files during \datamaintenance phase runs (\Cref{fig:w1_new_files_dm_1k}). 
Consequently, there are over $5x$ as many API calls to fetch the delta files for read queries (\Cref{fig:w1_storage_api_su_1k}). 
\Cref{fig:w1_perf_degradation} shows the ratio of runs of \singleuser and \datamaintenance phases for two scale factors, confirming the performance degradation observation for both scale factors and phase types. 
However, this behavior differs from Trino with Delta and Iceberg. 
\datamaintenance runs on Trino create up to $40x$ fewer files (\Cref{fig:w1_new_files_dm_1k}). 
We confirmed our hypothesis by running Spark \singleuser phase on Delta files created by Trino and observed the disappearance of previous degradation.

($ii$)~\emph{Trino demonstrates faster query execution compared to Spark.} 
During the build phase, both Spark and Trino generate an equal number of files. 
Consequently, in the initial \singleuser phase before any \datamaintenance phase, both Spark and Trino process the same number of files. 
However, our results in \Cref{fig:w1_1k_su_degradation} indicate that \singleuser~1 completes nearly twice as fast on Trino for both Delta and Iceberg. 
It is difficult to claim that Trino is consistently faster than Spark, as it depends on various factors like configuration, query type, plan, and data size. 
For instance, one contributing factor to Trino's speed advantage over Spark in our tests is its absence of checkpointing. 
While checkpointing enhances fault tolerance, it also introduces significant latency~\cite{starburst-trino-fault-tolerance}.

($iii$)~\emph{Spark with Hudi variation shows unmatched stability.} 
We now analyze the execution of \singleuser and \datamaintenance phases using Spark with Hudi (\Cref{fig:w1_perf_degradation,fig:w1_1k_su_degradation,fig:w1_1k_dm_degradation}). 
It exhibits distinct behavior compared to Delta and Iceberg, as the execution time, API calls, and data bytes read and written of the phases remains stable. 
Through our investigation, we discovered that Hudi has several default optimization-related parameters enabled, such as automatic cleanup and compaction~\cite{hudi-compaction-param}. 
Moreover, a crucial design choice in Hudi is to avoid creating small files by automatically adding sufficient records during the writing process to achieve the desired file size~\cite{hudi-sizing-param}. 
While these features contribute to stability in performance and efficiency, which is highly desirable, they come with trade-offs. 
We observe that, unlike Iceberg and Delta, Hudi-related variations read up to $\sim$6x more data and exhibit higher execution latencies. 
This observation is consistent with findings from other recent studies~\cite{lhbench}.

($iv$)~\emph{Read-Write tradeoff in MoR mode.} 
MoR optimizes frequent table updates, reducing data file rewriting costs and write latencies. 
However, it introduces a tradeoff: Increased computational cycles and network IO during read operations, impacting performance. 
To compare, we tested Hudi and Iceberg on Spark with CoW and MoR modes. 
\Cref{fig:w1_perf_1k_su,fig:w1_storage_api_su_1k} demonstrates that Iceberg and Hudi MoR versions consistently have slower performance than CoW variants due to the overheads mentioned earlier.

\begin{figure}
    \centering
    \includegraphics[width=\columnwidth]{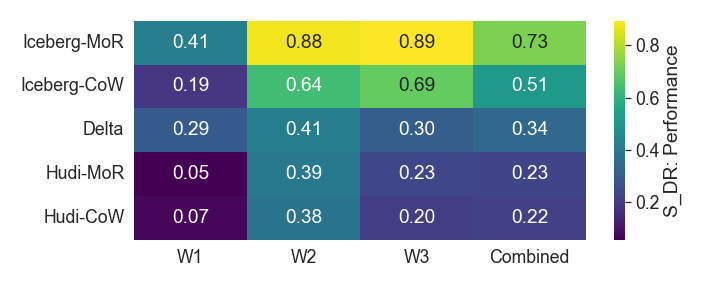}
    \caption{Performance Degradation, \(S_{DR}\), Evaluation. As lower \(S_{DR}\) is desired, Hudi emerges as most \emph{stable} \lst.}
    \label{fig:sdr_perf}
\end{figure}

\myparagraph{Performance Stability Analysis.}
To enable convenient stability comparisons despite the observed variability, we evaluate the effective performance degradation between two phases, \(S_{DR}\) (\S\ref{sec:measurements:stability}), for each experiment setup. The results are summarized in \Cref{fig:sdr_perf}. Each cell presents the combined performance \(S_{DR}\) value for \singleuser, \datamaintenance and \optimize phases across workloads \workload{1,2,3}, as well as the average for each setup. \Cref{fig:sdr_perf} is consistent with our prior discussion, confirming Hudi as the most stable \lst{}, particularly in read-intensive scenarios as indicated by \(S_{DR}\) below \(0.07\). Conversely, except in one instance against \workload{1}, Iceberg consistently exhibits lower stability, with a \(S_{DR}\) up to \(0.89\). Further analysis against \workload{2,3}, which involves optimization and concurrency, is presented in \S\ref{sec:eval:maintenance} and \S\ref{sec:eval:concurrency}.

\begin{figure*}
    \centering
    \begin{subfigure}[b]{0.65\textwidth}
        \centering
        \includegraphics[width=\textwidth]{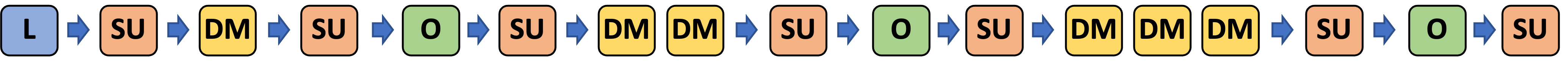}
    \end{subfigure}
    \begin{subfigure}[b]{0.65\textwidth}
        \centering
        \includegraphics[width=\textwidth]{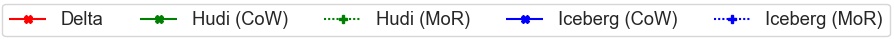}
    \end{subfigure}
    \hfill
    \begin{subfigure}[b]{\textwidth}
        \centering
        \includegraphics[width=.31\textwidth]{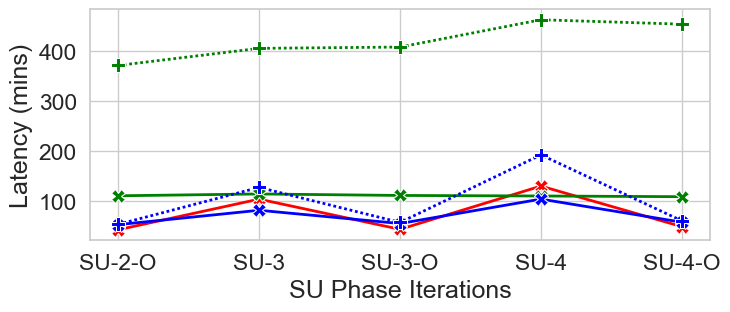}\quad
        \includegraphics[width=.31\textwidth]{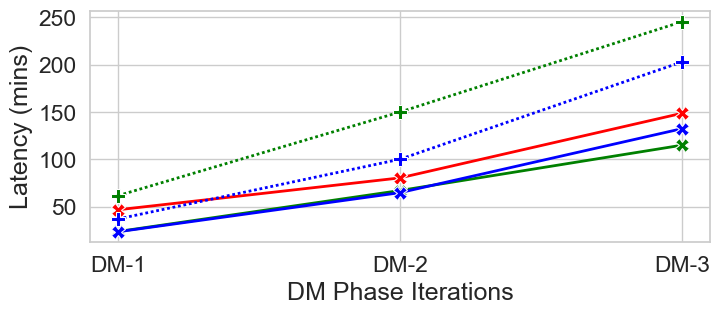}\quad
        \includegraphics[width=.31\textwidth]{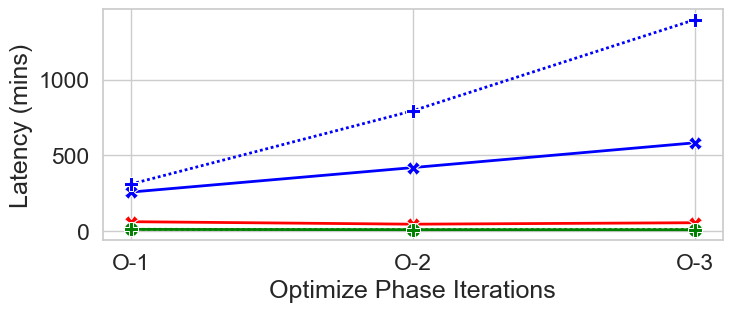}
        \caption{Phase Latency (in minutes). \singleuser performance recovers to pre-\datamaintenance levels due to \optimize.}
        \label{fig:eval:maintenance:latency_w2}
    \end{subfigure}
    \hfill
    \begin{subfigure}[b]{\textwidth}
        \centering
        \includegraphics[width=.31\textwidth]{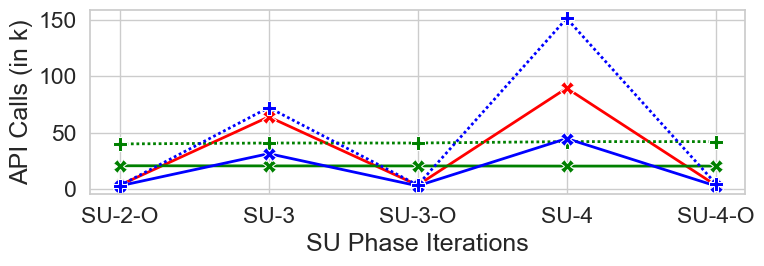}\quad
        \includegraphics[width=.31\textwidth]{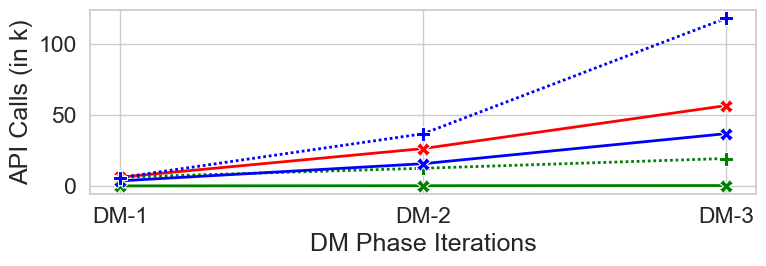}\quad
        \includegraphics[width=.31\textwidth]{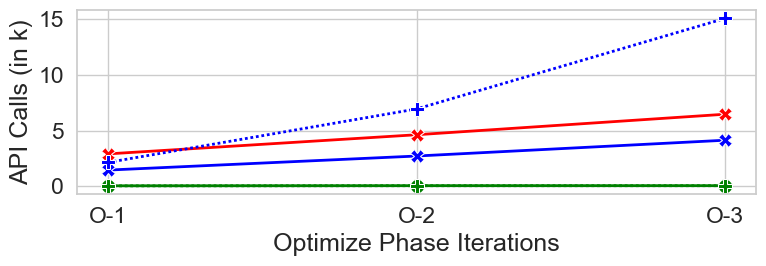}
        \caption{Storage API Calls (Millions). \singleuser results show significant reduction in API calls after \optimize.}
        \label{fig:eval:maintenance:calls_w2}
    \end{subfigure}
    \hfill
    \begin{subfigure}[b]{\textwidth}
        \centering
        \includegraphics[width=.31\textwidth]{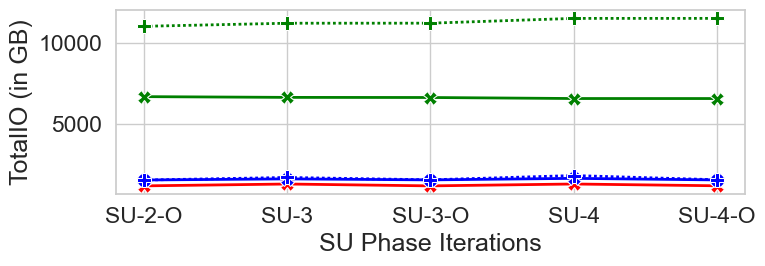}\quad
        \includegraphics[width=.31\textwidth]{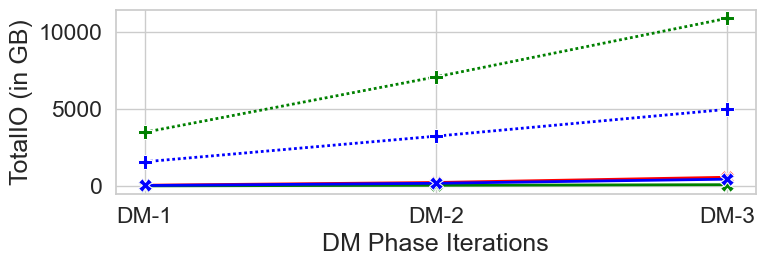}\quad
        \includegraphics[width=.31\textwidth]{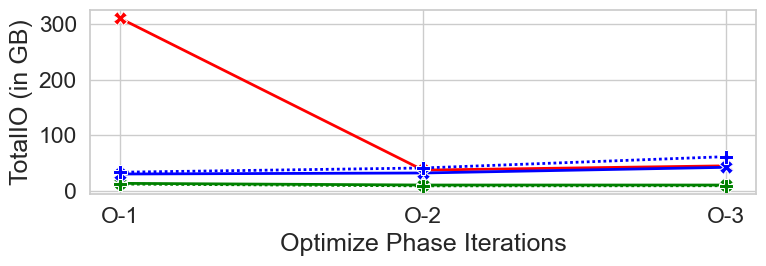}
        \caption{Total I/O Volume (in TB).}
        \label{fig:eval:maintenance:io_w2}
    \end{subfigure}
    \caption{Performance and storage efficiency evaluation of \workload{2} phases (SF1000).}
\end{figure*}

\subsection{Resilience}
\label{sec:eval:maintenance}
Next, we consider how the performance of \lsts changes when maintenance operations are introduced into the workload.
We use the \emph{resilience workload}~(\workload{2}, \S\ref{sec:benchmarks:tfe}), which evaluates the impact of the \optimize phase, i.e.,~compacting small data files into larger files for higher efficiency. 
The results for SF1000 are shown in \Cref{fig:eval:maintenance:latency_w2}.
\workload{2} iteratively executes a sequence of \singleuser (labeled SU-$i$), \datamaintenance (with an increasing number of tasks), \optimize, and \singleuser (labeled SU-$i$-O) tasks. Note for a given $i$, SU-$i$ and SU-$i$-O query the same logical version of the data. Results for SU-1 and SU-2 are not shown since they were discussed in \S\ref{sec:6.1_w1}.

We observe significantly different behavior for the three \lsts. 
Similar to the performance development in previous workloads, we observe that the addition of \optimize phases does not impact the performance of Hudi since its performance remains stable and only degrades minimally.
For both Iceberg and Delta, on the other hand, we observe that the \optimize phase has a significant impact on subsequent query execution of \singleuser phases, as shown by comparing SU-3 to SU-3-O and SU-4 to SU-4-O. 
We observe that the latency drops by 2.3x (1.5x for Iceberg-CoW and 2.2x for Iceberg-MoR) for the first pair and 2.6x (1.8x for Iceberg-CoW and 3.2x for Iceberg-MoR) for the second pair.
This indicates that periodic data maintenance operations are crucial for these \lsts to reduce the number of storage layer access calls (see \Cref{fig:eval:maintenance:calls_w2}) which are significantly higher in `unoptimized' phases.
For Hudi, we observe a high I/O volume in the \singleuser phase, solely composed of file read operations.
Furthermore, we see a relatively low amount of read or write activity for CoW in the \datamaintenance and \optimize phases (similar to Iceberg-CoW) while Hudi-MoR has a disproportionally high data volume in the \datamaintenance phase (similar to but more prominent than for Iceberg-MoR).

Interestingly, we observe a drastic increase in latency for Iceberg when executing \optimize phases, which cannot be observed for any other \lst.
Both I/O and CPU utilization remain comparatively low during these phases but by looking at the storage access calls in detail, we observe that Iceberg issues a large number of (sequentially executed) storage layer access calls.
The reason is that for each table, Iceberg's data compaction operation is executed file group-by-group by default~\cite{iceberg-compaction-group-param}. 
The operation parameters can be adjusted to run in parallel for $n$ groups, which can result in a significant performance boost, but making such tuning adjustments requires additional use case context.
Another interesting observation when looking at data maintenance cycles is a spike in I/O cost for the first \optimize phase executed by Delta.
This suggests that the operation makes significant changes to the data layout of the table generated after \load, reducing latency and I/O cost for subsequent data maintenance operations.

Finally, we observe that for this particular workload, CoW-based outperform MoR-based execution models, i.e.,~they have a lower latency and incur lower I/O cost.

\subsection{Read/Write Concurrency}
\label{sec:eval:concurrency}

\begin{figure}[t]
    \centering
    \begin{subfigure}[b]{0.75\columnwidth}
        \centering
        \includegraphics[width=\textwidth]{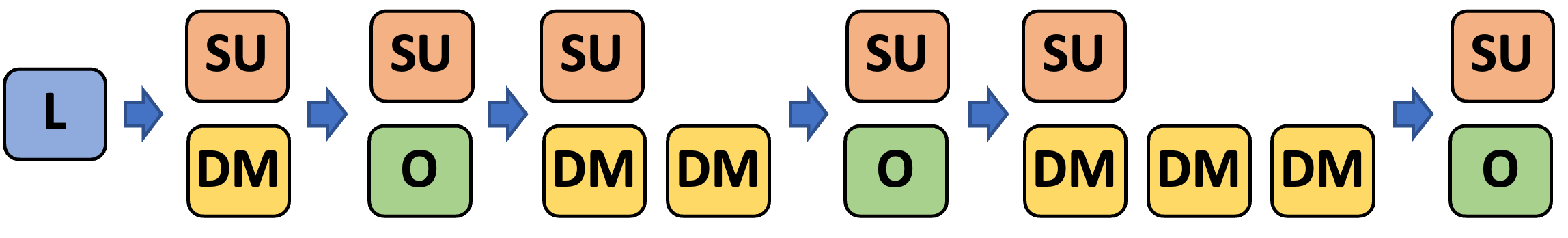}
    \end{subfigure}
    \hfill
    \begin{subfigure}[b]{\columnwidth}
    	\centering
	    \includegraphics[width=\columnwidth]{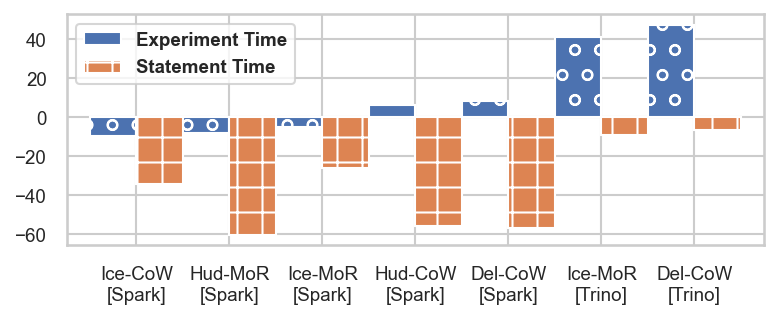}
	    \caption{Single cluster (\workload{3}).}
        \label{fig:w3_concurrency_perf_impact:single}
    \end{subfigure}
    \hfill
    \begin{subfigure}[b]{\columnwidth}
    	\centering
        \includegraphics[height=.14\paperheight,trim={0.22em 0 0.30em 0},clip]{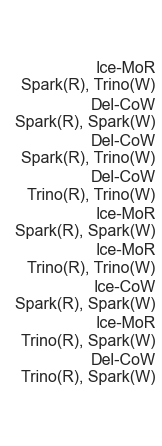}\hspace{0em}
        \includegraphics[height=.14\paperheight,trim={0.22em 0 0.7em 0},clip]{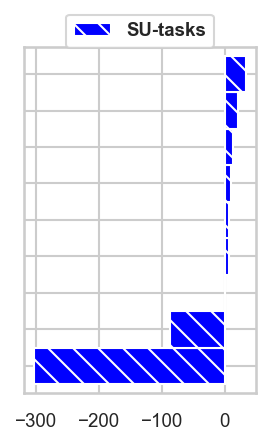}\hspace{-0.1em}
        \includegraphics[height=.14\paperheight,trim={0.16em 0 0.67em 0},clip]{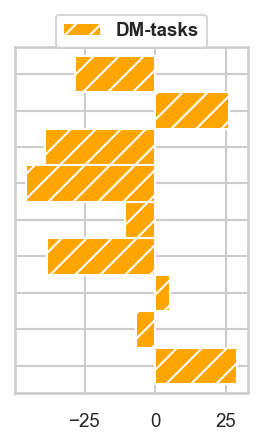}\hspace{-0.1em}
        \includegraphics[height=.14\paperheight,trim={0.16em 0 0.7em 0},clip]{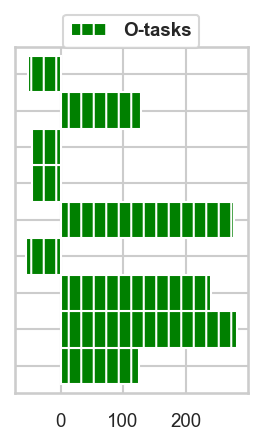}\hspace{-0.1em}
        \caption{Breakdown per task type on multiple clusters (\workload{3-Multi}). (R)~represents the reading engine (SU tasks), (W)~the writing engine (DM and O tasks).}
        \label{fig:w3_concurrency_perf_impact:multi}
    \end{subfigure}
    \caption{Performance gains and losses (in \%) of \workload{3} and \workload{3-Multi} relative to baseline \workload{2} (SF1000). Higher positive values indicate better performance.}
    \label{fig:w3_concurrency_perf_impact}
\end{figure}

Thus far, the sessions have been executed sequentially, utilizing all available resources. 
However, \lsts are designed with concurrency in mind. 
In this section, we analyze the impact of running read and write sessions concurrently using \workload{3}. 
Note that both \workload{2} and \workload{3} contain the same set and sequence of tasks. 
The only difference is that the phases of \workload{3} execute \singleuser task concurrently with either \datamaintenance or \optimize. 
Thus, we use \workload{2} as the baseline to evaluate execution speedup and overheads. 
The concurrent sessions can be executed either on a single cluster or on multiple clusters. 
In the latter setup, the \singleuser task is executed on our larger cluster, while the \datamaintenance and \optimize tasks are executed on our smaller one. 
We refer to this setup as \workload{3-Multi}\footnote{Hudi excluded since we encountered issues where queries failed due to updated underlying data, which prevented us from executing the \datamaintenance and \singleuser tasks concurrently on separate clusters.}. 

\Cref{fig:w3_concurrency_perf_impact:single} shows the comparison of the total execution time of the individual statements within all \singleuser, \datamaintenance, and \optimize tasks, referred to as \emph{Statement Time}, as well as the total end-to-end experiment execution time, which we call \emph{Experiment Time}, for a single cluster. 
In Spark, we observe that the \emph{Experiment Time} for \workload{3} is within a margin of $10\%$ compared to \workload{2}, while \emph{Statement Time} degrades by at least $25\%$ across all \lsts. 
Consequently, the concurrent execution of sessions in this setup does not necessarily lead to significant performance improvements due to resource contention. 
In contrast, Trino demonstrates more efficient utilization of the cluster resources, resulting in gains of at least $40\%$ in \emph{Experiment Time}, with only minor degradation observed in \emph{Statement Time}. 

We now turn our attention to the breakdown of \emph{Statement Time} per task type for the multiple cluster setup, depicted in \Cref{fig:w3_concurrency_perf_impact:multi}. 
To begin, when Spark executes \datamaintenance and \optimize tasks, Trino execution of \singleuser experiences a substantial slowdown compared to \workload{2}. 
Conversely, Spark execution of \singleuser shows a counterintuitive speed-up relative to \workload{2}. 
These differences can be attributed to the data layout generated by \datamaintenance tasks. 
For instance, using a smaller Spark cluster for \datamaintenance results in fewer generated files compared to a larger Spark cluster, though it still generates more files than Trino, as we discussed in \S\ref{sec:6.1_w1}. 
In all scenarios, reduced file reads in the multi-cluster setup leads to a decrease in task execution time. 
In relation to this observation, \optimize tasks in Spark are notably faster compared to \workload{2} due to fewer file reads, while \optimize tasks in Trino tend to be slower, primarily due to constrained resources in the smaller cluster. 
In summary, these results highlight the significant impact of setup and configuration values for \lsts, table maintenance operations, and engines on the overall performance, and the importance of considering real-world scenarios that previous benchmarks may have overlooked, such as the use of different engine combinations, in which \sys can help.

\subsection{Time Travel}

\begin{figure}
    \centering
    \includegraphics[width=\columnwidth]{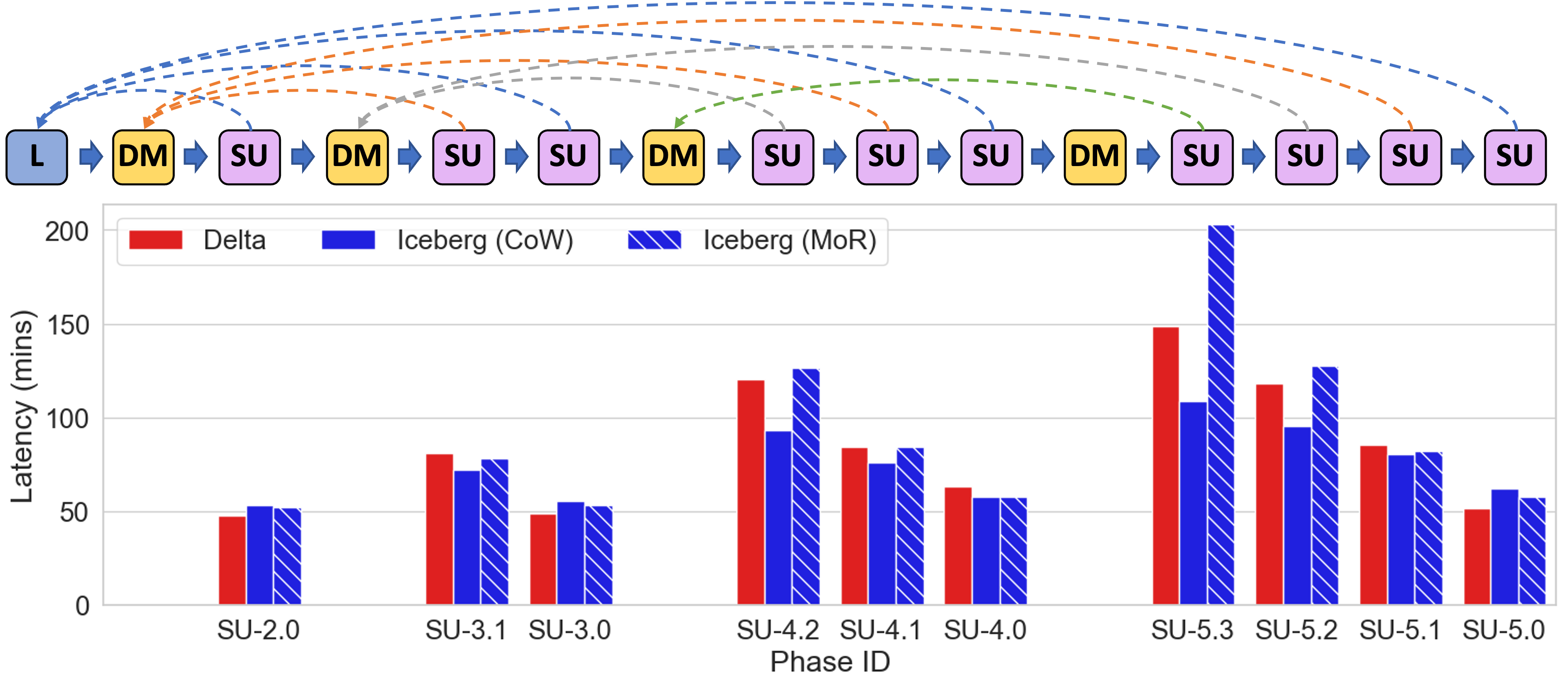}
    \caption{Latency measurements for \workload{4} (SF1000).}
    \label{fig:eval:tt:latency}
\end{figure}

Lastly, we study the performance of time travel queries in Spark\footnote{Hudi excluded due to SQL extension bug for time travel queries: \href{https://issues.apache.org/jira/browse/HUDI-7274}{HUDI-7274}.}, depicted in \Cref{fig:eval:tt:latency} for SF1000. 
In the figure, we use the notation SU-$i$.$v$ to represent the execution of a \singleuser task $i$ that queries version $v$ of the table. Here, $v=0$ corresponds to the table after the \load phase, while $v=j$ (where $j>0$) corresponds to the table after $j$ iteration of \datamaintenance.
The results are consistent with our expectations. 
We observe that the query latency increases as additional data files are written into the tables. 
This latency increase is similar to the increase observed in \workload{1} as new \datamaintenance operations are executed. In other words, writing new data after a specific version has been created appears to have minimal impact on querying that specific version. 
In addition, we analyze the storage efficiency (e.g. I/O and call counts) and find no significant difference between queries run on the latest version of the data and corresponding time travel queries on that version after data modifications have been applied. 
This indicates that these \lsts can efficiently support time travel queries without incurring any significant overhead in query performance or storage.

\section{Discussion}

\sys aims to serve various personas with distinct interests and requirements related to \lsts and their engine integrations, both existing and future ones. 
Notably, even in scenarios where access to infrastructure telemetry data is restricted, various personas can still leverage \sys by using custom patterns tailored to their specific workloads and objectives. 

($i$)~\emph{Developers from Engine-Building Organizations.} 
A primary user persona targeted by \sys is developers from organizations that build and commercialize platforms and query engines relying on \lsts. 
For instance, \company has recently adopted \sys for evaluating their system deployments, benefiting from valuable insights gained through these evaluations. 
They leverage \sys to compare various engine and \lst version combinations, and have plans to automate this process for tracking progress over time. 
Additionally, some of these organizations have recently focused on the development of a metadata translation layer from one \lst to another~\cite{delta-uniform,hudi-onetable}. 
In such cases, developers can use \sys to assess the effectiveness of the conversion process. 

($ii$)~\emph{Developers from Engine-Using Organizations.} 
Another key user persona includes developers from organizations that manage their own engines, even if they do not build them. 
Additionally, prospective customers seeking to compare and choose across platforms for a variety of workloads fall into this category. 
Some of these users have expressed interest in constructing their own \emph{benchmark packages} tailored to their specific scenarios, not necessarily building upon TPC-DS as the base workload. 
They can then employ \sys to perform various tasks, such as engine selection and configuration optimization. 

($iii$)~\emph{Researchers and Data Professionals.} 
\sys extends its utility to researchers and data professionals interested in studying best practices and tuning configurations across the three-dimensional space outlined in \S\ref{sec:otfs:overview}. 
For example, \sys can help to automate the selection of the optimal configuration for a table between CoW and MoR modes, which result in different trade-offs between read and write performance as demonstrated in \S\ref{sec:evaluation}. 
By expanding the variety of datasets and packages in \sys, these users can not only fine-tune configuration values for \lsts, table maintenance operations, or engines, but also investigate the existence of ``no-regret'' defaults that apply to these configuration parameters. 
Comparisons with the best defaults can be quantified to provide a clear understanding of the differences. 
While such evaluations are enabled by the \sys framework, they are beyond the scope of this paper. 

\myparagraph{Community Engagement and Adoption.} 
The decision to open-source \sys invites participation from organizations relying on \lsts for data processing. 
The engagement with these organizations has led to discussions about extensions to the tool in various areas, with some already integrated into the code.

($i$)~\emph{Workload Representation Model.} 
The existing model lacks a mechanism to define tasks \emph{dynamically} based on the data stored in \lsts, a feature valuable in various scenarios. 
For instance, one could create tasks to optimize a partitioned table, with each task executed on a user-configurable number of partitions. 
In another scenario, one could split the data maintenance task into queries that affect smaller data batches, each composed of a user-configurable number of rows. 
However, these patterns do not fit easily into the workload representation framework described in \S\ref{sec:benchmarks} because the number of tasks to execute is not known in advance. 
To address this challenge, we introduced the concept of \emph{parameterized custom tasks}, which expand the framework's capabilities by enabling the integration of custom user code for generating workflows dynamically. 

($ii$)~\emph{Libraries of Workload Components.} 
As explained in \S\ref{sec:framework}, \emph{tasks} in \sys are organized into libraries for convenient reuse across different workloads. 
Feedback from users indicated that this approach can lead to many redundant entries within workload pattern definition files. 
Consequently, we are investigating expanding our library model to incorporate definitions for other workload components, such as a \emph{session} or a \emph{phase}, which could then be shared and reused across multiple workloads as well.

($iii$)~\emph{Workload Packages.} 
Users recommended expanding the scope to include other standard benchmarks, like \mbox{TPC-H}, that are commonly used to evaluate analytical systems, as well as additional scenarios in which \lsts are commonly used, such as data cleaning~\cite{iceberg-trino-cleaning} and Change Data Capture (CDC) table mirroring with transactional consistency guarantees~\cite{iceberg-cdc}.

($iv$)~\emph{Metrics.} 
In the context of the CDC scenario, which allows various workflow designs (potentially dependent on the \lst capabilities), the choice of appropriate metrics for evaluating the scenario is still unclear. 
For example, users may prioritize \emph{data availability delay} over \emph{end-to-end execution time}.

\section{Related Work}
\label{sec:related}

The research and methods of evaluation of open \lsts are rapidly evolving, with new insights being published on a monthly basis. 
The bulk of literature is in the form of blog posts by vendors or users and is based on established benchmarks that were originally intended for a different category of systems. In this section, we first focus on works that compare \lsts both theoretically and empirically, then we discuss benchmarking methodology that is relevant for \lsts, and lastly, we discuss frameworks proposed in previous work.

\smallsection{Comparing \lsts}
Blog posts and papers focusing on the comparison of \lsts can be split into two categories. 
The first category is a theoretical evaluation of the different approaches, looking at features such as transaction management, schema evolution, and time travel~\cite{armbrust2021lakehouse, belov2021analysis, lhbench}.
Blog posts also commonly mention open-source statistics such as number of committers, pull requests, and Github ratings~\cite{lakehouse-feature-comparison, comparison-formats}.
After examining features and statistics, these blog posts often endorse one \lst over others.
In contrast, our objective is to establish a standardized approach to evaluate the performance and stability of \lsts, and to offer a framework that allows for empirical comparisons rather than just theoretical discussions.

The second category comprises comparisons of \lsts based on experimental evaluations.
For instance, in a recent paper, Jain et al.~\cite{lhbench} used TPC-DS benchmarking strategies adapted for \lsts to identify the strengths and weaknesses of each of them: ($i$)~They evaluate the impact of data updates on performance by running five sample queries, then merging changes into a table multiple times, and running the same queries again, and ($ii$)~they create a synthetic micro-benchmark with varying data refresh sizes to test the impact of the update size on the performance of the \lst.
Similarly, other recent blog posts~\cite{transparent-benchmarks,reassessing-performance} have also used the \load task (which has been modified to use Parquet as the source format~\cite{delta-benchmark}) and the \singleuser task from TPC-DS to compare performance of \lsts.

Drawing conclusions about the superiority of a specific \lst from the aforementioned works is challenging.
For instance, \cite{lhbench} found that ($i$)~Delta outperforms both Iceberg and Hudi in TPC-DS query performance, ($ii$)~Delta and Iceberg have significantly faster load times than Hudi, and ($iii$)~Delta provides better query performance after data has been modified than either Iceberg or Hudi.
Other works~\cite{transparent-benchmarks, reassessing-performance} suggest that Hudi is competitive with Delta in these same dimensions.
However, this does not mean that the results are misrepresented by any of those works, as engine setup and configuration parameters can significantly impact these evaluations~\cite{10.1145/3448016.3457569,DBLP:conf/icde/LinZFLZL22}.
Additionally, most evaluations use different versions of the \lsts and underlying execution engines further complicating objective comparisons.
To address these issues, we proposed an evaluation methodology and framework that is open source, customizable, repeatable, and easy to use, providing users with a one-stop solution to evaluate these formats objectively. 

\smallsection{Benchmarking \lsts}
Prior research has used various benchmarks to compare \lsts, but the most commonly used benchmark is TPC-DS.
It is important to note that while TPC-DS~V1~\cite{DBLP:conf/vldb/OthayothP06} was designed to evaluate monolithic RDBMSs, TPC-DS~V2~\cite{DBLP:conf/cloud/PoessRJ17} was specifically developed to cater to SQL-based big data systems. 
However, even though TPC-DS~V2 already considered SQL engines running on a common storage layer (e.g., HDFS) that could be accessed by multiple systems, it ignored key elements in evaluating \lsts, such as data layout optimization, time travel, or the impact of data manipulations over extended time periods.
Furthermore, the TPC-DS result consists of a single performance metric, which, although straightforward for ranking purposes, is insufficient in capturing critical \lst-based concepts like \emph{stability}~(\S\ref{sec:measurements}).
We have therefore proposed metrics that complement the TPC-DS performance score and can help to evaluate a system across these additional dimensions.

As mentioned above, prior work~\cite{lhbench} has taken a first step towards modifying TPC-DS by evaluating the performance difference of a set of queries before and after a series of SQL {\small\textrm{MERGE INTO}} statements were executed.
With our work, we take the idea of long-term impact evaluation one step further and make TPC-DS composable, i.e.,~we allow users to create their own workloads based on TPC-DS by mixing and matching the different TPC-DS phases.
This modification allows us to evaluate all of the previously unaddressed elements that are unique to \lsts.

Prior work has also examined customized micro-benchmarks designed to evaluate the read and write capabilities of different \lsts. For instance, these benchmarks include operations that append and remove data from an existing table, and mimic GDPR deletions~\cite{brooklyn-data}. 
Integrating these workloads into our benchmarking framework to further extend the evaluation should be a straightforward process.

\smallsection{Benchmarking Frameworks} %
It is important to note that previous research has developed several benchmarking frameworks mainly geared towards evaluating SQL systems.
For example, OLTPBench~\cite{oltpbench} can execute several standardized workloads using different database backends such as PostgreSQL or SQL Server.
Similarly, DIAMetrics~\cite{diametrics} was designed to allow its users to compare and contrast the execution of different (customizable) benchmarks, also extending the idea of benchmarking to include other aspects such as data movement and data security.
In this paper, we describe a framework that is specifically focused on \lsts.
In addition to their specific SQL dialects, \lsts may be executed on different types of clusters, with different optimization parameters, for which we deploy different (and novel) benchmarks in their evaluation.
DSB~\cite{dsb} focuses on workload-driven RDBMSs that adapt over time using ML techniques, extending TPC-DS with more complex data distribution, query templates, and dynamic workloads. 
In contrast, \sys concentrates on \lsts but could easily integrate DSB's modifications to expand the range of evaluated scenarios. 

YCSB~\cite{ycsb} presents a framework to compare cloud data \emph{serving} systems like Cassandra or HBase using tiers and workloads. 
YCSB and \sys share similarities in their approach, but YCSB focuses solely on this category of systems, and does not cover aspects such as the composability of workloads, which is a key contribution of our work. 
Finally, PEEL~\cite{peel} is designed for benchmarking distributed systems, with a focus on reproducibility and automated experiment processes. 
Our implementation also takes inspiration from PEEL, although we recognize the need for further automation in \sys to run evaluations more effectively in cloud deployments, which we plan to explore in future work. 

\section{Conclusion}

In this paper, we presented \sys, our benchmarking framework for evaluation of \lsts using workload patterns that mimic real-world customer scenarios, such as those found within \company, while at the same time providing means to fairly evaluate those workloads. 
We discuss in-depth how we enable users to create their own custom workloads using the extendable plug and execute functionality within \sys and showcase how we use \sys to evaluate and compare existing \lsts.
Our extensive evaluation finds that these \lsts vary significantly in terms of their performance, storage efficiency, and stability.
This demonstrates that \sys can be used to evaluate \lsts effectively and comprehensively.

\begin{acks}
We would like to thank Joyce Cahoon and Yiwen Zhu for their valuable discussions on the stability metric, and Jose Medrano for his feedback on the benchmarking framework. 
Additionally, we appreciate the early users and individuals who have shown interest in \sys, as their feedback is helping us steer the project in the right direction.
\end{acks}

\balance

\bibliographystyle{ACM-Reference-Format}
\bibliography{ref}

\end{document}